\documentclass[journal]{IEEEtran}
\IEEEoverridecommandlockouts
\usepackage{cite}
\usepackage{amsmath,amssymb,amsfonts}
\usepackage{algorithmic}
\usepackage{graphicx}
\usepackage{textcomp}
\usepackage{xcolor}
\usepackage{xpatch}
\usepackage{amsfonts}
\usepackage{bm}
\usepackage{booktabs}
\usepackage{algorithm}
\usepackage{multirow,bm,bbm,array,setspace}
\usepackage{textcomp}
\usepackage{pstricks,enumerate}
\usepackage{bbm,subfigure}

\makeatletter
\newcommand{\distas}[1]{\mathbin{\overset{#1}{\kern\z@\sim}}}%
\newsavebox{\mybox}\newsavebox{\mysim}
\newcommand{\distras}[1]{%
  \savebox{\mybox}{\hbox{\kern1pt$\scriptstyle#1$\kern1pt}}%
  \savebox{\mysim}{\hbox{$\sim$}}%
  \mathbin{\overset{#1}{\kern\z@\resizebox{\wd\mybox}{\ht\mysim}{$\sim$}}}%
}
\ExplSyntaxOn
\cs_new:Npn \bibColoredItems #1#2
  {
    \clist_map_inline:nn {#2} { \cs_new:cpn {bib@colored@##1} {#1} } 
  }
\ExplSyntaxOff

\newcommand\bib@setcolor[1]{%
  \ifcsname bib@colored@#1\endcsname
    \expanded{\noexpand\color{\csname bib@colored@#1\endcsname}}%
  \else
    \normalcolor
  \fi
}

\IfPackageLoadedTF{hyperref}{\@tempswatrue}{\@tempswafalse}
\if@tempswa
  \xpatchcmd\@bibitem {\H@item}{\bib@setcolor{#1}\H@item}{}{\PatchFailed}
  \xpatchcmd\@lbibitem{\H@item}{\bib@setcolor{#2}\H@item}{}{\PatchFailed}
\else
  \xpatchcmd\@bibitem {\item}  {\bib@setcolor{#1}\item}  {}{\PatchFailed}
  \xpatchcmd\@lbibitem{\item}  {\bib@setcolor{#2}\item}  {}{\PatchFailed}
\makeatother
\ifCLASSINFOpdf
\else
\fi

\newcommand{\bH}{\bm{H}}

\newcommand{\bg}{\bm{g}}

\newcommand{\bs}{\bm{s}}
\newcommand{\bI}{\bm{I}}

\newcommand{\bv}{\bm v}

\newcommand{\bY}{\bm{Y}}
\newcommand{\by}{\bm{y}}
\newcommand{\bz}{\bm{z}}
\newcommand{\bZ}{\bm{Z}}
\newcommand{\bQ}{\bm{Q}}
\newcommand{\bx}{\bm{x}}
\newcommand{\bA}{\bm{A}}
\newcommand{\ba}{\bm{a}}
\newcommand{\bB}{\bm{B}}
\newcommand{\bG}{\bm{G}}
\newcommand{\bM}{\bm{M}}
\newcommand{\bF}{\bm{F}}
\newcommand{\bX}{\bm{X}}

\newcommand{\bK}{\bm{K}}

\newcommand{\bL}{\bm{L}}
\newcommand{\bD}{\bm{D}}

\newcommand{\hh}{\mathrm{H}}

\newtheorem{proposition}{Proposition}
\newtheorem{lemma}{Lemma}

\newtheorem{remark}{Remark}

\begin{document}

\title{Accelerating Quadratic Transform and WMMSE}
\author{
\IEEEauthorblockN{
Kaiming Shen, \IEEEmembership{Senior Member,~IEEE}, Ziping Zhao, \IEEEmembership{Member,~IEEE}, Yannan Chen, \IEEEmembership{Student Member,~IEEE},\\ Zepeng Zhang, \IEEEmembership{Student Member,~IEEE}, and Hei Victor Cheng, \IEEEmembership{Member,~IEEE}
} 
\thanks{Manuscript accepted to IEEE Journal on Selected Areas in Communications in May 2024. The work of Kaiming Shen and Yannan Chen was supported in part by the National Natural Science Foundation of China (NSFC) under Grant 92167202 and in part by Guangdong Major Project of Basic and Applied Basic Research (No. 2023B0303000001). The work of Hei Victor Cheng was supported in part by the Aarhus Universitets Forskningsfond under Project AUFF 39001. \emph{(Corresponding author: Kaiming Shen.)}

Kaiming Shen and Yannan Chen are with the School of Science and Engineering, The Chinese University of Hong Kong (Shenzhen), 518172 Shenzhen, China (e-mail: shenkaiming@cuhk.edu.cn; yannanchen@link.cuhk.edu.cn).

Ziping Zhao and Zepeng Zhang are with the School of Information Science and Technology, ShanghaiTech University, 201210 Shanghai, China (e-mail: zipingzhao@shanghaitech.edu.cn; zhangzp1@shanghaitech.edu.cn).

Hei Victor Cheng is with the Electrical and Computer Engineering Department, Aarhus Universrity, 8200 Aarhus, Denmark (e-mail: hvc@ieee.org).
}
}

\maketitle

\begin{abstract}
Fractional programming (FP) arises in various communications and signal processing problems because several key quantities in the field are fractionally structured, e.g., the Cram\'{e}r-Rao bound, the Fisher information, and the signal-to-interference-plus-noise ratio (SINR). A recently proposed method called the quadratic transform has been applied to the FP problems extensively. The main contributions of the present paper are two-fold. First, we investigate how fast the quadratic transform converges. To the best of our knowledge, this is the first work that analyzes the convergence rate for the quadratic transform as well as its special case the weighted minimum mean square error (WMMSE) algorithm. Second, we accelerate the existing quadratic transform via a novel use of Nesterov's extrapolation scheme \cite{Nesterov_book}. Specifically, by generalizing the minorization-maximization (MM) approach in \cite{ZP_MM+}, we establish a nontrivial connection between the quadratic transform and the gradient projection, thereby further incorporating the gradient extrapolation into the quadratic transform to make it converge more rapidly. Moreover, the paper showcases the practical use of the accelerated quadratic transform with two frontier wireless applications: integrated sensing and communications (ISAC) and massive multiple-input multiple-output (MIMO).
\end{abstract}

\begin{IEEEkeywords}
Fractional programming (FP), convergence rate, acceleration, minorizatioin-maximization (MM).
\end{IEEEkeywords}

\section{Introduction}

\IEEEPARstart{F}{ractional} programming (FP) is a study of optimization aimed at ratio terms. 
With the matrix coefficients $\bA, \bB_1, \ldots,\bB_n$ and the vector variables $\bx_1,\ldots,\bx_n$, this paper focuses on the following type of ratio term:
\begin{align*}
\big(\bA\bx_i\big)^\hh\Bigg(\sum^n_{j=1}\bB_{j}\bx_{j}\bx_{j}^\hh\bB_{j}^\hh\Bigg)^{-1}\big(\bA\bx_i\big),
\end{align*}
where $(\cdot)^\hh$ stands for the conjugate transpose. A further matrix generalization of the scalar-valued ratio is also considered in the paper. The above ratio is of significant research interest not only because it is a natural extension of the Rayleigh quotient, but also because many key quantities in communications and signal processing are written in this form, e.g., the Cram\'{e}r-Rao bound, the Fisher information, and the signal-to-interference-plus-noise ratio (SINR).

\subsection{Motivation}

The quadratic transform \cite{shen2018fractional1,shen2019optimization} is a state-of-the-art tool for FP. Differing from the classic FP methods, i.e., the Charness-Cooper method \cite{charnes1962programming} and Dinkelbach's method \cite{dinkelbach1967nonlinear}, that are limited to the single-ratio case, the quadratic transform is capable of decoupling the numerator and denominator for multiple ratios simultaneously and thus renders the multi-ratio FP problem easier to address. The quadratic transform has wide connecting links to other optimization methods, e.g., \cite{shen2018fractional1} shows that the quadratic transform boils down to a fixed-point iteration method when applied to the power control problem, while \cite{shen2019optimization} interprets the quadratic transform as a minorization-maximization (MM) method \cite{razaviyayn2013unified,sun2016majorization} in general. In particular, \cite{shen2018fractional2} shows that the quadratic transform encompasses the weighted minimum mean square error (WMMSE) algorithm for the beamforming problem \cite{cioffi_WMMSE,Shi_WMMSE} as a special case. Furthermore, \cite{shen2018fractional2} proposes a better way to use the quadratic transform than WMMSE for the discrete FP problems. 

Although the theory and applications of the quadratic transform have been considerably developed over the past few years, the convergence rate of the quadratic transform remains a complete mystery (which is unknown even for the special case of the quadratic transform, the WMMSE algorithm), with the following open problems:
\begin{itemize}
\it
    \item How fast does the quadratic transform converge?
    \item How is it compared to the conventional gradient method?
    \item Can we further accelerate the quadratic transform?
\end{itemize}
The analysis in this paper sheds light on all these questions. Roughly speaking, we show that: (i) if the starting point is sufficiently close to a strict local optimum, then the quadratic transform yields an error bound of $O(1/k)$, where $k$ is the number of iterations; (ii) the quadratic transform converges faster than the gradient method in iterations, but more slowly in time due to its higher per-iteration complexity; (iii) the proposed accelerated quadratic transform has a much lower per-iteration complexity and reduces the error bound to $O(1/k^2)$.

\subsection{Literature Review}

The traditional study on FP focuses on the ratio between two scalar-valued functions \cite{stancu2012fractional,YaFeng_survey}, the simplest case of which is the single-ratio problem. 
Despite the nonconvexity, the single-ratio problem can be efficiently solved either by the Charnes-Cooper method 
\cite{charnes1962programming} or by Dinkelbach's method \cite{dinkelbach1967nonlinear} so long as the concave-convex condition \cite{shen2018fractional1} holds. In contrast, the multi-ratio FP is much more challenging. Except for the max-min-ratios case that can be efficiently solved by a generalized Dinkelbach's method \cite{crouzeix1985algorithm}, it is difficult to extend the classic FP technique to the multi-ratio problems. It turns out that even the sum-of-ratios problem is NP-complete \cite{freund2001solving} so that the existing approaches mostly resort to the branch-and-bound algorithm \cite{freund2001solving,thi2003unified,konno2000branch,benson2001global,qu2007efficient,kuno2002branch,liu2019new,benson2007solving,benson2002global,benson2002using} and thus incur exponential complexities. As such, the early applications of FP in communications and signal processing are restricted to the energy efficiency problem \cite{zappone2015energy}, namely the maximization of a single ratio of energy efficiency. The advent of the quadratic transform \cite{shen2018fractional1} has significantly extended the scope of the applications of FP by enabling efficient multi-ratio optimization. Aside from the sum-of-ratios problem, the quadratic transform works for (or can be extended to) the log-ratio problem \cite{shen2018fractional2}, the matrix-ratio problem \cite{shen2019optimization}, and the mixed max-and-min FP problem \cite{max_min_FP}.

As a special case of the quadratic transform, the WMMSE algorithm \cite{cioffi_WMMSE, Shi_WMMSE} has been extensively studied for its own sake because of the critical role it plays in the beamforming design for multiple-input multiple-output (MIMO) wireless networks.
The WMMSE algorithm entails computing the matrix inverse on a per-iteration basis---which is costly in modern and future wireless networks because the matrix size grows with the number of antennas. For this reason, the computational complexity has long been identified as a major bottleneck of the WMMSE algorithm. Assuming that the channel matrices are all full row-rank, the recent work \cite{Shi_rethink_2023} 
exploits the weighted sum-rate (WSR) problem structure to facilitate the matrix inverse operation. The more recent work in \cite{Zhao_ICASSP2023,ZP_MM+} goes further. It does not require any channel assumptions and can sidestep the matrix inverse operation in all cases. In particular, \cite{ZP_MM+} shows that its proposed beamforming algorithm without matrix inversion boils down to a gradient projection. We will show that the above results of \cite{ZP_MM+} for the WSR problem carry over to a broad range of other FP problems. Moreover, another recent endeavor \cite{ZL_ICASSP2023} suggests combining Nesterov's extrapolation and WMMSE in a heuristic fashion, but the resulting algorithm still incurs the matrix inverse operation and cannot give any performance guarantees.

\subsection{Main Results}

The main results of this paper divide into three parts. First, we propose two new FP methods with provable higher efficiency than the existing quadratic transform. By the nonhomogeneous bounding technique from the MM theory \cite{sun2016majorization}, we devise the \emph{nonhomogeneous quadratic transform} to eliminate the matrix inverse operation. Geometric insight is provided into the advantage of the nonhomogeneous quadratic transform over the existing quadratic transform. Further, based on the connection with the gradient projection, we incorporate Nesterov's extrapolation to accelerate convergence, namely the \emph{extrapolated quadratic transform}.

Of equal importance in this work is the convergence rate analysis for the quadratic transform, which has never been considered in the literature to date. Under some mild condition that renders the nonconvex analysis tractable, we show that the existing quadratic transform (which encompasses WMMSE \cite{cioffi_WMMSE,Shi_WMMSE} as a special case) yields faster convergence in iterations than the gradient projection and the nonhomogeneous quadratic transform, but possibly slower convergence in time, and yields an error bound of $O(1/k)$ after $k$ iterations. In comparison, the extrapolated quadratic transform leads to a superior error bound of $O(1/k^2)$.

The last part of the main results in this paper concerns the practical applications of the accelerated quadratic transform. Because the new FP technique is free of matrix inversion and has fast convergence, it is particularly suited for those frontier wireless technologies involving large antenna arrays. We present two application cases: (i) the integrated sensing and communications (ISAC) and (ii) the massive MIMO, neither of which can be efficiently handled by the existing quadratic transform \cite{cioffi_WMMSE, Shi_WMMSE} due to the curse of dimensionality.


\subsection{Paper Organization and Notation}

The remainder of the paper is organized as follows. Section \ref{sec:prob} describes the sum-of-weighted-ratios FP problem. Section \ref{sec:QT} connects the quadratic transform \cite{shen2018fractional1} to gradient projection, thus devising its acceleration based on Nesterov's extrapolation scheme. Section \ref{sec:convergence} analyzes the convergence rates for the different quadratic transform methods. Section \ref{sec:other_MFP} discusses other FP problems. Section \ref{sec:matrix_var} discusses the extension to the matrix ratio case. Two application cases are presented in Section \ref{sec:applications}. Finally, Section \ref{sec:conclusion} concludes the paper.

Here and throughout, bold lower-case letters represent vectors while bold upper-case letters represent matrices. For a vector $\ba$, $\ba^c$ is its complex conjugate, $\ba^\hh$ is its conjugate transpose, and $\|\ba\|_2$ is its $\ell_2$ norm.
For a matrix $\bA$, $\bA^c$ is its complex conjugate, $\bA^\top$ is its transpose, $\bA^\hh$ is its conjugate transpose, $\lambda_{\max}(\bA)$ is its largest eigenvalue, $\|\bA\|_F$ is its Frobenius norm, and $\mathrm{tr}(A)$ is its trace. For a positive semi-definite matrix $\bA$, $\bA^\frac{1}{2}$ is its square root. Denote by $\bI$ the identity matrix, $\mathbb C^{\ell}$ the set of $\ell\times1$ vectors, $\mathbb C^{d\times m}$ the set of $d\times m$ matrices, and $\mathbb S_{++}^{d\times d}$ the set of $d\times d$ positive definite matrices. For a complex number $a\in\mathbb C$, $\Re\{a\}$ is its real part. An underlined letter represents a collection of variables denoted by the same letter, e.g., $\underline\ba = \{\ba_1, \ba_2, \ldots, \ba_n\}$.


\section{Problem Statement}
\label{sec:prob}

Consider a total of $n$ ratio terms, each written as
\begin{align}
\label{matrix-ratio}
 M_i(\underline\bx) = \big(\bA_i\bx_i\big)^\hh\Bigg(\sum^n_{j=1}\bB_{ij}\bx_{j}\bx_{j}^\hh\bB_{ij}^\hh\Bigg)^{-1}\big(\bA_i\bx_i\big)
\end{align}
for $i=1,\ldots,n$, where $\bA_i\in\mathbb C^{\ell\times d},\bB_{ij}\in\mathbb C^{\ell\times d}$, and $\bx_j\in\mathbb C^d$. Let $f_o(\underline\bx)$ be a positive linear combination of these ratios:
\begin{align}
\label{fo}
f_o(\underline\bx) = \sum^n_{i=1} \omega_iM_i(\underline\bx),
\end{align}
where each weight $\omega_i>0$. We consider the following sum-of-weighted-ratios FP problem:
\begin{subequations}
\label{prob:MFP}
\begin{align}
  \underset{\underline\bx}{\text{maximize}} &\quad f_o(\underline\bx)\\
  \text{subject to} & \quad \bx_i\in\mathcal X_i,\quad i=1,\ldots,n,
\end{align}
\end{subequations}
where $\mathcal X_i$ is a nonempty convex constraint set on $\bx_i$. Accordingly, the Cartesian product $\mathcal{X}=\mathcal X_1\times\mathcal X_2\times\ldots\times\mathcal X_n$ is the constraint set on $\underline\bx$.

Notice that we can insert constant terms into the numerator and denominator of $M_i(\underline\bx)$ by using the dummy variable $\bx_j=[1,\ldots,1]^\top$. 
Further, when each vector variable $\bx_i\in\mathbb C^d$ is generalized to the matrix form $\bX_i\in\mathbb C^{d\times m}$, the ratio term also becomes a matrix:
\begin{align}
\label{matrix:ratio}
\bM_i(\underline\bX) = \big(\bA_i\bX_i\big)^\hh\Bigg(\sum^n_{j=1}\bB_{ij}\bX_j\bX_j^\hh\bB_{ij}^\hh\Bigg)^{-1}\big(\bA_i\bX_i\big).
\end{align}
The above matrix-FP case is dealt with in Section \ref{sec:matrix_var}. We further consider other FP problem types in Section \ref{sec:other_MFP}.

\section{Quadratic Transform}
\label{sec:QT}

We start by reviewing the quadratic transform \cite{shen2018fractional1,shen2018fractional2}, then establish its connection to the gradient projection, and lastly exploit this connection to incorporate Nesterov's extrapolation scheme \cite{Nesterov_book} into the quadratic transform and thereby accelerate the convergence.

\subsection{Preliminary}

The main idea of the quadratic transform is to decouple the numerator and denominator for each ratio term. The following proposition specializes the quadratic transform \cite{shen2018fractional1} to our problem case in \eqref{prob:MFP}.
\begin{proposition}
\label{prop:QT}
The sum-of-weighted-ratios FP problem \eqref{prob:MFP}
is equivalent to
\begin{subequations}
\label{QT:new_prob}
\begin{align}
\underset{\underline\bx,\,\underline\by}{\text{maximize}} &\quad f_q(\underline\bx,\underline\by)\\
  \text{subject to} &\quad \bx_i\in\mathcal X_i,\;\by_i\in\mathbb C^\ell,\quad i=1,\ldots,n
\end{align}
\end{subequations}
with the new objective function
\begin{equation}
\label{fq:form1}
f_q(\underline\bx,\underline\by)
    = \sum^n_{i=1}\omega_i\Bigg[2\Re\{\bx^\hh_i\bA^\hh_i\by_i\}-\sum^n_{j=1}\by^\hh_i\bB_{ij}\bx_j\bx^\hh_j\bB^\hh_{ij}\by_i\Bigg],
\end{equation}
in the sense that $\underline\bx^\star$ is a solution of problem \eqref{prob:MFP} if and only if $(\underline\bx^\star,\underline\by^\star)$ is a solution of problem \eqref{QT:new_prob} with some optimal auxiliary variable $\underline\by^\star$.
\end{proposition}

We consider optimizing $\underline\bx$ and $\underline\by$ alternatingly in the new problem. When $\underline\bx$ is held fixed, by completing the square for each $\by_i$ in \eqref{fq:form1}, the optimal $\by_i$ can be obtained as
\begin{equation}
\label{QT:y}
    \by_i^\star = \Bigg(\sum^n_{j=1}\bB_{ij}\bx_j\bx^\hh_j\bB^\hh_{ij}\Bigg)^{-1}\big(\bA_i\bx_i\big).
\end{equation}
To solve for $\underline\bx$ with $\underline\by$ held fixed, we regroup the terms in \eqref{fq:form1} according to $\bx_i$ and thus rewrite $f_q(\underline\bx,\underline\by)$ as
\begin{equation}
\label{QT:fq}
   f_q(\underline\bx,\underline\by) = \sum^n_{i=1}\Big[2\Re\{\omega_i\bx^\hh_i\bA^\hh_i\by_i\}-\bx^\hh_i\bD_i\bx_i\Big],
\end{equation}
where 
\begin{equation}
\label{Di}
    \bD_i = \sum^n_{j=1}\omega_j\bB_{ji}^\hh\by_j\by^\hh_j\bB_{ji}.
\end{equation}
Each $\bx_i$ can now be optimally determined as
\begin{equation}
\label{QT:x}
    \bx^\star_i = \arg\min_{\bx_i\in\mathcal X_i}\big\|\bD^{\frac12}_i\big(\bx_i-\omega_i\bD_i^{-1}\bA^\hh_i\by_i\big)\big\|_2.
\end{equation}
Geometrically, the optimal solution requires projecting the vector $\omega_i\bD_i^{-1}\bA^\hh_i\by_i$ onto the set $\mathcal X_i$ under the linear transformation $\bD^{\frac12}_i$.
In particular, if $\omega_i\bD_i^{-1}\bA^\hh_i\by_i\in\mathcal X_i$ already, then the optimal solution reduces to
\begin{equation}
\label{QT:x simple}
    \bx_i^\star = \omega_i\bD_i^{-1}\bA^\hh_i\by_i.
\end{equation}
Alternating between step \eqref{QT:y} and step \eqref{QT:x} constitutes the conventional quadratic transform method, as summarized in Algorithm \ref{algorithm:QT}.

\begin{algorithm}[t]
  \caption{Conventional Quadratic Transform \cite{shen2018fractional1}}
  \label{algorithm:QT}
  \begin{algorithmic}[1]
      \STATE Initialize $\underline\bx$ to a feasible value.
      \REPEAT 
      \STATE Update each $\by_i$ according to \eqref{QT:y}.
      \STATE Update each $\bx_i$ according to \eqref{QT:x}.
      \UNTIL{the value of $f_o(\underline\bx)$ converges} 
  \end{algorithmic}
\end{algorithm}

Although $\underline\bx$ can now be iteratively updated in closed form by Algorithm \ref{algorithm:QT}, the matrix inversion in step \eqref{QT:x} becomes quite costly when $\bD_i$ is a large matrix---which is the case for the applications involving large antenna arrays as discussed in Section \ref{sec:applications}.



\subsection{Connection with Gradient Projection}
\label{subsec:gradient}

The following result generalizes the WSR maximization algorithm in \cite{ZP_MM+}. We first introduce two lemmas.
\begin{lemma}
\label{lemma:derivative}
After $\underline\by$ has been updated as in \eqref{QT:y} for the current $\underline\bx$,
the partial derivative of each fractional function $M_i(\underline\bx)$ with respect to the complex conjugate\footnote{The motivation of adopting $\partial M_i(\underline\bx)/\partial \bx^c_j$ rather than $\partial M_i(\underline\bx)/\partial \bx_j$ is that the corresponding differential is simpler. According to Theorem 2 in \cite{hjorungnes2007complex}, the two types of partial derivatives are both feasible for computing the stationary point. In the rest of the paper, we shall always use the former type.} of $\bx_j$ is given by
\begin{equation*}
\frac{\partial M_i(\underline\bx)}{\partial\bx^c_j}= 
\left\{ 
\begin{array}{ll}
    \!\!\bA^\hh_i\by_i-\bB_{ii}^\hh\by_i\by_i^\hh\bB_{ii}\bx_i & \text{if}\;j=i
    \vspace{0.5em}\\
    \!\!-\bB_{j i}^\hh\by_j\by_j^\hh\bB_{j i}\bx_i & \text{otherwise}.\\
\end{array}
\right.
\end{equation*}
\end{lemma}

\begin{lemma}[Nonhomogeneous Bound \cite{sun2016majorization}]
\label{lemma:Taylor}
Suppose that the two Hermitian matrices $\bL,\bK\in\mathbb C^{d\times d}$ satisfy $\bL\preceq\bK$. Then for any two vectors $\bx,\bz\in\mathbb C^d$, one has
\begin{equation}
\label{Taylor_inequality}
\bx^\hh\bL\bx\le\bx^\hh\bK\bx+2\Re\{\bx^\hh(\bL-\bK)\bz\}+\bz^\hh(\bK-\bL)\bz,
\end{equation}
where the equality holds if $\bz=\bx$. This bound is so called due to the nonhomogeneous term $2\Re\{\bx^\hh(\bL-\bK)\bz\}$.
\end{lemma}

Treating $\bD_i$ as $\bL$ in \eqref{Taylor_inequality}, we let
\begin{equation}
\label{lambda}
    \bK = \lambda_{i}\bI,\quad\text{where}\;\;\lambda_i\ge\lambda_{\max}(\bD_i)
\end{equation}
so that $\bL\preceq\bK$; one possible choice is $\lambda_i=\|\bD_i\|_F$. Thus, by Lemma \ref{lemma:Taylor}, we can bound $f_q(\underline\bx,\underline\by)$ in \eqref{QT:fq} from below as 
\begin{equation}
\label{fq<ft}
    f_q(\underline\bx,\underline\by) \ge f_t(\underline\bx,\underline\by,\underline\bz)
\end{equation}
with
\begin{multline}  
\label{ft}
f_t(\underline\bx,\underline\by,\underline\bz) = \sum^n_{i=1} \Big[ 2\Re\big\{\omega_i\bx^\hh_i\bA^\hh_i\by_i+\bx^\hh_i(\lambda_i\bI-\bD_i)\bz_i\big\}\\
    +\bz^\hh_i(\bD_i-\lambda_i\bI)\bz_i-\lambda_i\bx^\hh_i\bx_i\Big].
\end{multline}
In particular, the equality in \eqref{fq<ft} holds if $\bz_i=\bx_i$ for all $i$.

We then use the lower bound $f_t(\underline\bx,\underline\by,\underline\bz)$ to approximate the new objective function $f_q(\underline\bx,\underline\by)$. Observe that the three variables $(\underline\bx,\underline\by,\underline\bz)$ can be iteratively optimized in closed form in $f_t(\underline\bx,\underline\by,\underline\bz)$.
When $\underline\by$ and $\underline\bx$ are both held fixed, the optimal update of $\underline\bz$ follows by the equality condition in Lemma \ref{lemma:Taylor} as
\begin{equation}
\label{GQT:z}
    \bz_i^\star = \bx_i,\;\;\text{for}\;i=1,\ldots,n.
\end{equation}
After $\underline\bz$ has been updated to $\underline\bx$, the optimal $\by_i$ in \eqref{ft} is still determined as in \eqref{QT:y}.
Next, when $\underline\by$ and $\underline\bz$ are both held fixed, the optimal $\bx_i$ in \eqref{ft} is given by
\begin{align}
    \bx_i^\star &= \arg\min_{\bx_i\in\mathcal X_i}\big\|\lambda_i\bx_i-\omega_i\bA^\hh_i\by_i-\big(\lambda_i\bI-\bD_i\big)\bz_i\big\|_2\notag\\
    &= \mathcal{P}_{\mathcal X_i}\bigg(\bz_i + \frac{1}{\lambda_i}\big(\omega_i\bA^\hh_i\by_i-\bD_i\bz_i\big)\bigg),
    \label{GQT:x}
\end{align}
where $\mathcal P_{\mathcal X_i}(\cdot)$ is the Euclidean projection on $\mathcal X_i$. Algorithm \ref{algorithm:GQT} summarizes the above steps.

\begin{figure}[t]
\centering
\includegraphics[width=1.0\linewidth]{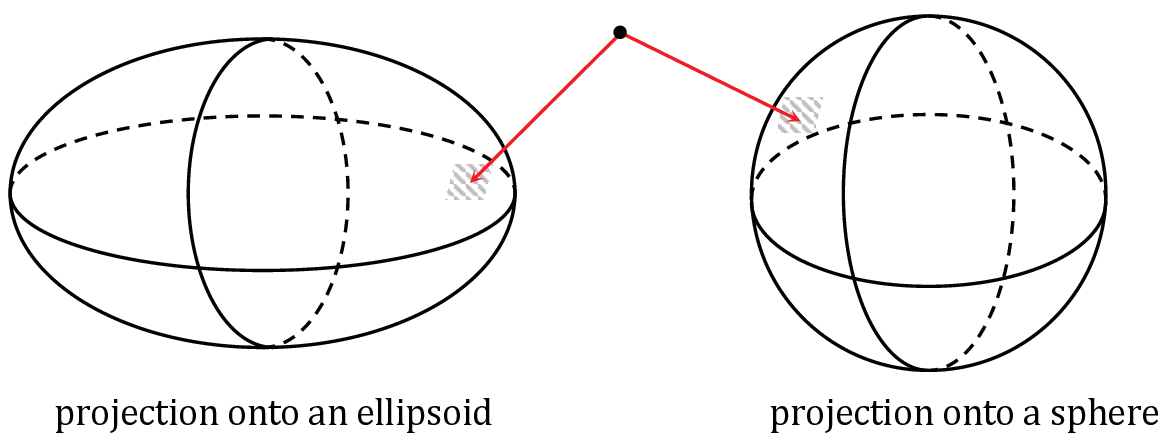}
\caption{{The conventional quadratic transform amounts to the projection onto an ellipsoid and incurs matrix inverse operation. In contrast, the new quadratic transform avoids matrix inverse by computing the projection onto a sphere.}}
\label{fig:projection}
\end{figure}

\begin{figure}[t]
\centering
\includegraphics[width=8.8cm]{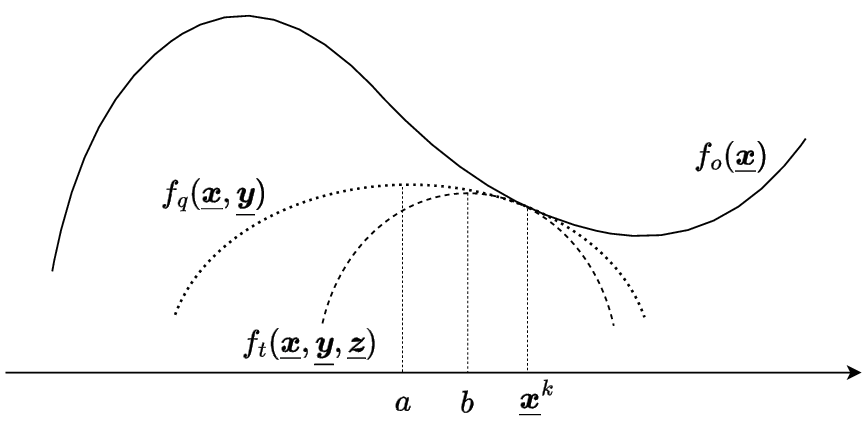}
\caption{Algorithm \ref{algorithm:QT} approximates $f_o(\underline\bx)$ as $f_q(\underline\bx,\underline\by)$ while Algorithm \ref{algorithm:GQT} approximates $f_o(\underline\bx)$ as $f_t(\underline\bx,\underline\by,\underline\bz)$. By the MM procedure, for the current solution $\underline\bx^k$, Algorithm \ref{algorithm:QT} updates it to $a$, while Algorithm \ref{algorithm:GQT} updates it to $b$. Algorithm \ref{algorithm:QT} converges faster in iterations because its approximation is tighter.}
\label{fig:MM}
\end{figure}

We are now ready to interpret the above iterative optimization as a gradient projection method. We use the superscript $k=1,2,\ldots$ to index the iteration, and assume that the three variables $(\underline\bx,\underline\by,\underline\bz)$ are cyclically updated as
\begin{equation*}
\underline\bx^0\rightarrow\cdots\rightarrow\underline\bx^{k-1} \rightarrow \underline\bz^k \rightarrow \underline\by^k \rightarrow \underline\bx^k \rightarrow \underline\bz^{k+1} \rightarrow \cdots.
\end{equation*}
With the optimal $\underline\by$ in \eqref{QT:y} and the optimal $\underline\bz$ in \eqref{GQT:z} substituted into \eqref{GQT:x}, the optimal update of $\bx_i$ in iteration $k$ can be recognized as
\allowdisplaybreaks
\begin{align}
    \bx^{k}_i &= \mathcal{P}_{\mathcal X_i}\bigg(\bz^k_i + \frac{1}{\lambda^{k}_i}\big(\omega_i\bA^\hh_i\by^k_i-\bD^k_i\bz^k_i\big)\bigg)\notag\\
    &\overset{(a)}{=} \mathcal{P}_{\mathcal X_i}\bigg(\bx^{k-1}_i + \frac{1}{\lambda^{k}_i}\big(\omega_i\bA^\hh_i\by^k_i-\bD^k_i\bx^{k-1}_i\big)\bigg)\notag\\
    &\overset{(b)}{=} \mathcal{P}_{\mathcal X_i}\Bigg(\bx^{k-1}_i + \frac{1}{\lambda^{k}_i}\sum^n_{j=1}\Bigg[\omega_j\cdot\frac{\partial  M_j(\underline\bx^{k-1})}{\partial \bx^c_i}\Bigg]\Bigg)\notag\\
    &= \mathcal{P}_{\mathcal X_i}\bigg(\bx^{k-1}_i + \frac{1}{\lambda^{k}_i}\cdot\frac{\partial f_o(\underline\bx^{k-1})}{\partial \bx^c_i}\bigg),
    \notag
\end{align}
\interdisplaylinepenalty=10000
in which $\bD_i$ is assigned the iteration index $k$ because it has been updated by \eqref{Di} for $\underline\by^{k}$, and $\lambda_i$ is assigned the iteration index $k$ because it is impacted by $\bD^k_i$. Here, step $(a)$ follows by \eqref{GQT:z}, and step $(b)$ follows by Lemma \ref{lemma:derivative}.

\begin{algorithm}[t]
  \caption{Nonhomogeneous Quadratic Transform}
  \label{algorithm:GQT}
  \begin{algorithmic}[1]
      \STATE Initialize $\underline\bx$ to a feasible value.
      \REPEAT 
      \STATE Update each $\bz_i$ according to \eqref{GQT:z}.
      \STATE Update each $\by_i$ according to \eqref{QT:y}.
      \STATE Update each $\bx_i$ according to \eqref{GQT:x}.
      \UNTIL{the value of $f_o(\underline\bx)$ converges} 
  \end{algorithmic}
\end{algorithm}

\begin{remark}[Geometric Interpretation]
The solution \eqref{QT:x} in Algorithm \ref{algorithm:QT} by the conventional quadratic transform amounts to the projection onto an ellipsoid, while the solution \eqref{GQT:x} in Algorithm \ref{algorithm:GQT} by the nonhomogeneous quadratic transform amounts to the projection onto a sphere, as shown in Fig.~\ref{fig:projection}. The sphere projection is computationally much easier.
\end{remark}

\begin{remark}
The connection of the quadratic transform to the gradient projection was initially proposed in \cite{ZP_MM+} for the WSR problem (which is a log-ratio type of FP problem), as specified in Section \ref{sec:other_MFP}. This paper extends this connection considerably.
\end{remark}

\begin{figure*}[t]
\centering
\subfigure[convergence in iterations when $n=5$, $d=9$, and $\ell=4$]{
\includegraphics[width=0.45\linewidth]{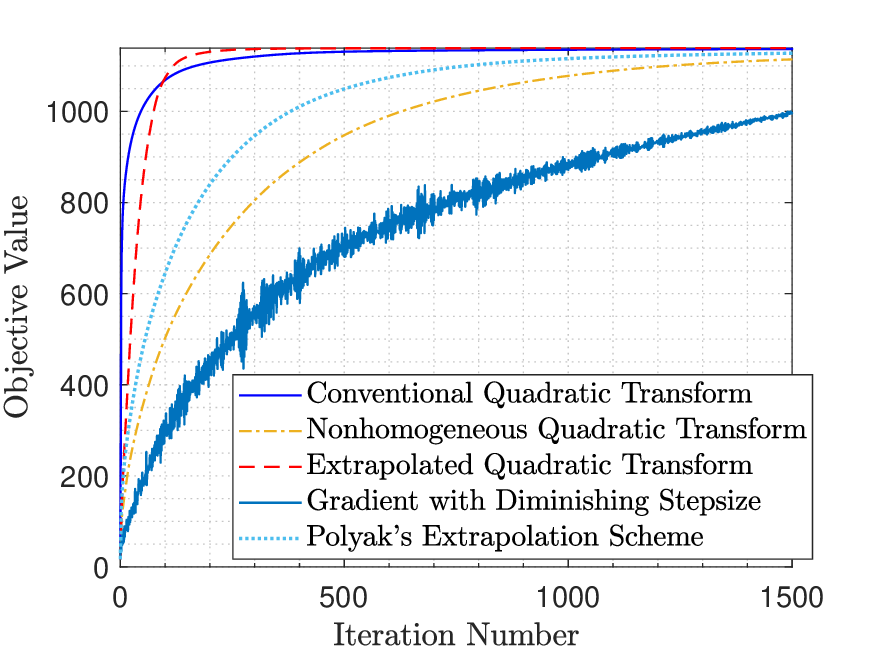}}
\hspace{1em}
\subfigure[convergence in time when $n=5$, $d=9$, and $\ell=4$]{
\includegraphics[width=0.45\linewidth]{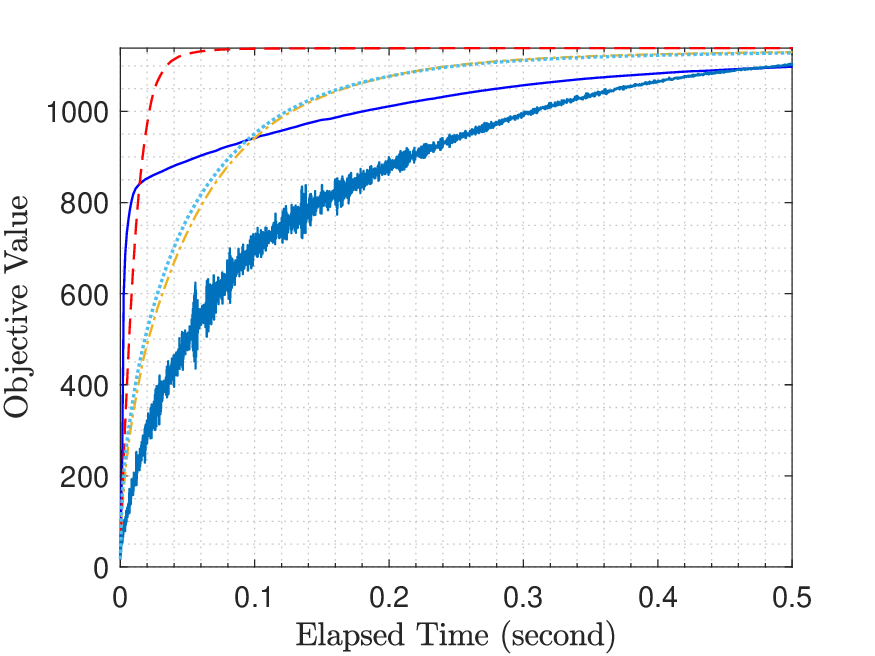}}\\
\subfigure[convergence in iterations when $n=5$, $d=20$, and $\ell=10$]{
\includegraphics[width=0.45\linewidth]{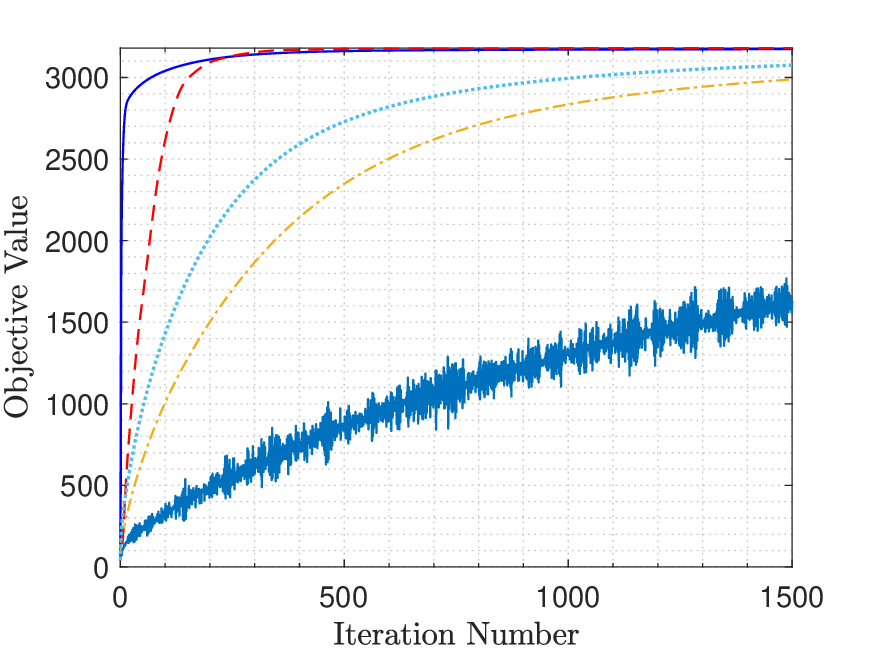}}
\hspace{1em}
\subfigure[convergence in time when $n=5$, $d=20$, and $\ell=10$]{
\includegraphics[width=0.45\linewidth]{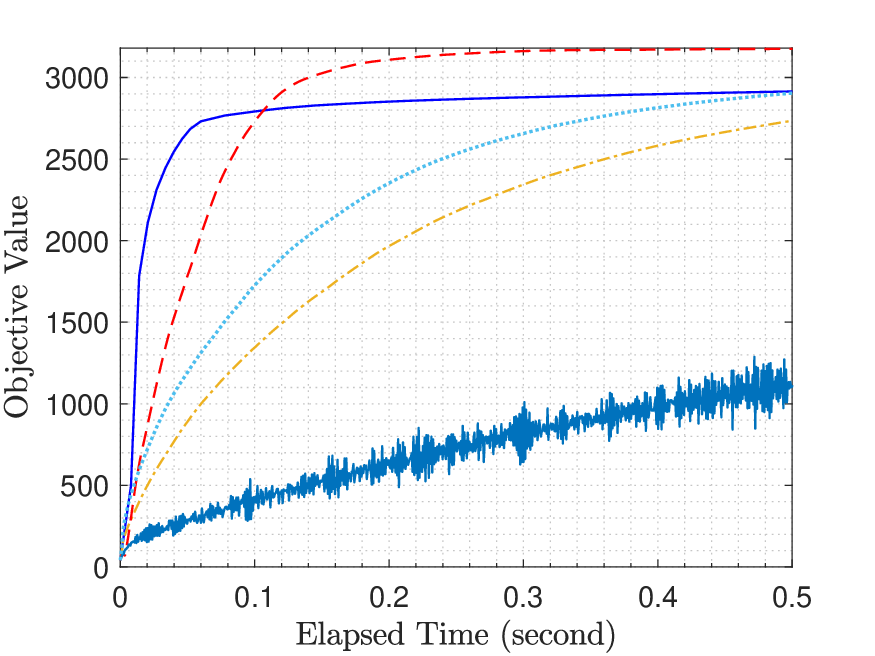}}
\caption{Average performance of solving 100 random examples of problem \eqref{prob:MFP}. Let each $\omega_i=1$, let each $\mathcal X_i=\{\bX\in\mathbb C^{d\times\ell}:\text{tr}(\bX\bX^\hh)\le10\}$, and randomly generate each entry of $\bA_i$ and $\bB_{ij}$ i.i.d. according to $\mathcal{CN}(0,1)$; further, add $\bI$ to each matrix denominator to ensure its positive definiteness.}
\label{fig:convergence}
\end{figure*}

\subsection{Accelerated Quadratic Transform}
\label{subsec:AQT}


In terms of the per-iteration complexity, it is evident that Algorithm \ref{algorithm:GQT} is more efficient since it does not\footnote{Notice that Algorithm \ref{algorithm:GQT} still requires computing the $\ell\times\ell$ matrix inverse when updating $\underline\by$. Actually, this matrix inverse can be also eliminated by applying Lemma \ref{lemma:Taylor} one more time, and consequently it would introduce a new group of auxiliary variables. We do not consider this straightforward extension in this paper because $\ell$ is quite small in our application cases and thus it is unnecessary to eliminate the matrix inverse in \eqref{QT:y}.} require computing matrix inverse for the iterative update of $\underline\bx$. The overall complexity is however much more difficult to examine because it also depends on how many iterations the algorithm entails to reach the convergence. Algorithm \ref{algorithm:QT} uses $f_q(\underline\bx,\underline\by)$ to approximate the original objective $f_o(\underline\bx)$ from below, while Algorithm \ref{algorithm:GQT} further uses $f_t(\underline\bx,\underline\by,\underline\bz)$ to approximate $f_q(\underline\bx,\underline\by)$ from below, i.e.,
\begin{equation*}
    f_o(\underline\bx)\ge f_q(\underline\bx,\underline\by)\ge f_t(\underline\bx,\underline\by,\underline\bz).
\end{equation*}
Intuitively, Algorithm \ref{algorithm:QT} has faster convergence than Algorithm \ref{algorithm:GQT} in iterations because it uses a tighter approximation of $f_o(\underline\bx)$, as illustrated in Fig.~\ref{fig:MM}. A formal analysis of their convergence rates is provided in Section \ref{sec:convergence}.

To sum up, Algorithm \ref{algorithm:GQT} is more efficient per iteration, but in the meanwhile requires more iterations to attain convergence. One can find a balance between Algorithm \ref{algorithm:QT} and Algorithm \ref{algorithm:GQT} via timesharing; the convergence to a stationary point is still guaranteed by the MM theory as shown in Section \ref{sec:convergence}.

{A brand-new idea here is to reduce the number of iterations for Algorithm \ref{algorithm:GQT} by means of extrapolation.} In principle, since the nonhomogeneous quadratic transform in essence utilizes the gradients, its convergence can be accelerated by momentum or heavy-ball method. Specifically, following Nesterov's extrapolation strategy \cite{Nesterov_book}, we propose to extrapolate each $\bx_i$ along the direction of the difference between the preceding two iterates before the gradient projection, i.e.,
\begin{align}
    \bm{\nu}^{k-1}_i &= \bx^{k-1}_i+\eta_{k-1}(\bx^{k-1}_i-\bx^{k-2}_i)
    \label{extrapolate}\\
    \bx^{k}_i &= \mathcal{P}_{\mathcal X_i}\bigg(\bm{\nu}_i^{k-1} + \frac{1}{\lambda^{k}_i}\cdot\frac{\partial f_o(\underline{\bm{\nu}}^{k-1})}{\partial \bx^c_i}\bigg),
\end{align}
where the extrapolation step $\eta_{k}$ can be chosen as
\begin{equation*}
    \eta_k = \max\bigg\{\frac{k-2}{k+1},0\bigg\},\quad\text{for}\;k\ge1,
\end{equation*}
and the starting point is $\bx^{-1}=\bx^0$. The above gradient projection with extrapolation can be implemented with the assistance of the auxiliary variables $(\underline\by,\underline\bz)$ as shown in Algorithm \ref{algorithm:EQT}, which is referred to as the extrapolated quadratic transform.


\begin{algorithm}[t]
  \caption{Extrapolated Quadratic Transform}
  \label{algorithm:EQT}
  \begin{algorithmic}[1]
      \STATE Initialize $\underline\bx$ to a feasible value.
      \REPEAT 
      \STATE Update each $\bm{\nu}_i$ according to \eqref{extrapolate} and set $\bx_i=\bm{\nu}_i$.
      \STATE Update each $\bz_i$ according to \eqref{GQT:z}.
      \STATE Update each $\by_i$ according to \eqref{QT:y}.
      \STATE Update each $\bx_i$ according to \eqref{GQT:x}.
      \UNTIL{the value of $f_o(\underline\bx)$ converges} 
  \end{algorithmic}
\end{algorithm}

Fig.~\ref{fig:convergence} compares the above three algorithms numerically. Aside from Algorithms 1 to 3, we consider the gradient method with a diminishing stepsize $1/k$ and also a variant of Algorithm \ref{algorithm:EQT} by using Polyak's extrapolation \cite{Poljak_extrapolation} instead of Nesterov's. Observe that Algorithm \ref{algorithm:QT} converges faster than Algorithm \ref{algorithm:EQT} in iterations according to Fig.~\ref{fig:convergence}(a) and Fig.~\ref{fig:convergence}(c), but more slowly in time as shown in Fig.~\ref{fig:convergence}(b) and Fig.~\ref{fig:convergence}(d).

\section{Convergence Analysis}
\label{sec:convergence}

We have introduced three types of quadratic transform: the conventional quadratic transform (Algorithm \ref{algorithm:QT}), the nonhomogeneous quadratic transform (Algorithm \ref{algorithm:GQT}), and the extrapolated quadratic transform (Algorithm \ref{algorithm:EQT}). This section comprises two main results. First, we show that the three types of quadratic transform all guarantee convergence to a stationary-point solution. Second, we analyze the rate of convergence for them.


The convergence proof uses the MM interpretation of the quadratic transform from \cite{shen2019optimization,ZP_MM+}. Write the optimal update of $\underline\by$ in \eqref{QT:y} as a function of $\underline\bx$:
\begin{equation*}
    \mathcal Y(\underline\bx) = \underline\by \;\, \text{with each} \;\, \by_i = \Bigg(\sum^n_{i=1}\bB_{ij}\bx_j\bx^\hh_j\bB^\hh_{ij}\Bigg)^{-1}\big(\bA_i\bx_i\big).
\end{equation*}
By Algorithm \ref{algorithm:QT}, after $\underline\by^k$ is optimally updated for the previous $\underline\bx^{k-1}$, the current new objective function $f_q(\underline\bx,\underline\by)$ can be rewritten as a function $r_q(\underline\bx|\underline\bx^{k-1})$ of $\underline\bx$ conditioned on $\underline\bx^{k-1}$:
\begin{equation}
\label{rq}
    r_q(\underline\bx|\underline\bx^{k-1}) = f_q(\underline\bx,\mathcal Y(\underline\bx^{k-1})),
\end{equation}
and accordingly the update of $\underline\bx$ in \eqref{QT:x} can be rewritten as
\begin{equation}
\label{MM:fq}
    \underline\bx^{k} = \arg\max_{\underline\bx\in\mathcal X} r_q(\underline\bx|\underline\bx^{k-1}).
\end{equation}
Importantly, it always holds that
\begin{equation*}
    r_q(\underline\bx|\underline\bx^{k-1}) \le f_o(\underline\bx)\;\;\text{and}\;\; r_q(\underline\bx^{k-1}|\underline\bx^{k-1}) = f_o(\underline\bx^{k-1}),
\end{equation*}
so updating $\underline\by$ for $\underline\bx^{k-1}$ is equivalent to constructing a surrogate function $r_q(\underline\bx|\underline\bx^{k-1})$ for $f_o(\underline\bx)$ at $\underline\bx^{k-1}$, namely the \emph{minorization} step. Moreover, \eqref{MM:fq} can be recognized as the \emph{maximization} step. As such, Algorithm \ref{algorithm:QT} boils down to an MM method, and hence it guarantees convergence to a stationary point of problem \eqref{prob:MFP}. By a similar argument, we can also interpret Algorithm \ref{algorithm:GQT} as an MM method, with the surrogate function
\begin{equation}
    r_t(\underline\bx|\underline\bx^{k-1}) = f_t(\underline\bx,\mathcal Y(\underline\bx^{k-1}),\underline\bx^{k-1}).
\end{equation}
Besides, the tradeoff between Algorithm \ref{algorithm:QT} and Algorithm \ref{algorithm:GQT} via timesharing constitutes an MM algorithm too and hence preserves the stationary-point convergence.
Furthermore, recall that Algorithm \ref{algorithm:GQT} can also be interpreted as a gradient projection method; since it has provable convergence to a stationary point, so does its accelerated version Algorithm \ref{algorithm:EQT}. The following proposition summarizes the above results based on the MM interpretation.

\begin{proposition}
\label{prop:MM}
For the sum-of-weighted-ratios FP problem \eqref{prob:MFP}, the conventional quadratic transform in Algorithm \ref{algorithm:QT}, the nonhomogeneous quadratic transform in Algorithm \ref{algorithm:GQT}, and the extrapolated quadratic transform in Algorithm \ref{algorithm:EQT} all guarantee convergence to some stationary point, with the original objective value increasing monotonically.
\end{proposition}

We further analyze the rate of convergence for the various quadratic transform methods. Due to the nonconvexity of the FP problems, the global analysis (assuming that the starting point is far from any stationary point) is intractable. We would like to give a local analysis by restricting the constraint set to a small neighborhood of a strict local optimum (so that the starting point is not far away), i.e.,
\begin{equation}
\label{small_constraint}
    \mathcal X = \big\{\underline\bx:\|\underline\bx-\underline\bx^*\|_2\le R\big\},
\end{equation}
where $\underline\bx^*$ is a strict local optimum of \eqref{prob:MFP} satisfying
\begin{equation*}
    \nabla^2 f_o(\underline\bx^*) \preceq -\xi\bI \prec \mathbf 0
\end{equation*}
for some strictly positive constant $\xi>0$, and the radius $R>0$ is sufficiently small so that $f_o(\underline\bx)$ is concave on $\mathcal X$. Assume that the Hessian of $f_o(\underline\bx)$ is $L$-Lipschitz continuous on $\mathcal X$, i.e.,
\begin{equation*}
    \|\nabla^2f_o(\underline\bx)-\nabla^2f_o(\underline\bx')\|_2\le L\|\underline\bx-\underline\bx'\|_2
\end{equation*}
for any $\underline\bx,\underline\bx'\in\mathcal X$. By Corollary 1.2.2 of \cite{Nesterov_book}, we have
\begin{equation*}
    \nabla^2f_o(\underline\bx)\preceq  \nabla^2f_o(\underline\bx^*) + L\|\underline\bx-\underline\bx^*\|_2\bI,
\end{equation*}
so it suffices to require $R\le \xi/L$ to ensure the concavity of $f_o(\underline\bx)$ on $\mathcal X$.

The convergence rate analysis also uses the MM interpretation. Conditioned on $\underline\bx'\in\mathcal X$, write the difference between $f_o(\underline\bx)$ and each surrogate function as a function of $\underline\bx\in\mathcal X$:
\begin{align*}
\delta_q(\underline\bx|\underline\bx') &= f_o(\underline\bx) - f_q(\underline\bx,\mathcal Y(\underline\bx'))\\
\delta_t(\underline\bx|\underline\bx') &= f_o(\underline\bx) - f_t(\underline\bx,\mathcal Y(\underline\bx'),\underline\bx').
\end{align*}
It can be readily shown that
\begin{subequations}
\label{delta:conditions}
\begin{align}
&\delta_q(\underline\bx^k|\underline\bx^{k}) = \delta_t(\underline\bx^k|\underline\bx^{k}) = 0  \\
&\nabla\delta_q(\underline\bx^k|\underline\bx^{k}) = \nabla\delta_t(\underline\bx^k|\underline\bx^{k}) = {0}.
\end{align}    
\end{subequations}
Moreover, define the following quantities to be the largest eigenvalues of two Hessian matrices:
\begin{align*}
\Lambda_q &= \max_{\underline\bx\in\mathcal X} \lambda_{\max}\big(\nabla^2\delta_q(\underline\bx|\underline\bx)\big)\\
\Lambda_t &= \max_{\underline\bx\in\mathcal X} \lambda_{\max}\big(\nabla^2\delta_t(\underline\bx|\underline\bx)\big).
\end{align*}
Recall each ratio $M_i(\underline\bx)$ is finite with a nonsingular denominator matrix $\sum^n_{i=1}\bB_{ij}\bx_j\bx^\hh_j\bB^\hh_{ij}$, so each entry of $\underline\by$ is finite and hence each $\lambda_{\max}(\bD_i)<\infty$ according to \eqref{Di}. As a result, $\lambda_{\max}(\nabla^2_{\underline\bx}f_t(\underline\bx,\underline\by,\underline\bz)=-\max_i2\lambda_{\max}(\bD_i)>-\infty$. Further, $ \Lambda_t\le \lambda_{\max}(\nabla^2_{\underline\bx}f_o(\underline\bx))-\lambda_{\max}(\nabla^2_{\underline\bx}f_t(\underline\bx,\mathcal Y(\underline\bx'),\underline\bx'))<\infty$. Moreover, because $\nabla^2_{\underline\bx}f_t(\underline\bx,\mathcal Y(\underline\bx'),\underline\bx')\preceq \nabla^2_{\underline\bx}f_o(\underline\bx,\mathcal Y(\underline\bx'))$, we must have $\Lambda_q\le\Lambda_t<\infty$. 
We are now ready to derive the (local) convergence rates of Algorithm \ref{algorithm:QT} and Algorithm \ref{algorithm:GQT}, as stated in the following proposition.

\begin{proposition}[Convergence Rates of Algorithm \ref{algorithm:QT} and Algorithm \ref{algorithm:GQT}]
\label{prop:MM_converge}
Suppose that the Hessian of the objective function $f_o(\underline\bx)$ of problem \eqref{prob:MFP} is $L$-Lipschitz continuous and that the radius of the constraint set $\mathcal X$ is sufficiently small as defined in \eqref{small_constraint}. Then the local convergence rate of Algorithm \ref{algorithm:QT} or Algorithm \ref{algorithm:GQT} is
\begin{align}
f_o(\underline\bx^*)-f_o(\underline\bx^{1}) &\le \frac{\Lambda R^2}{2}+\frac{LR^3}{6}\\
f_o(\underline\bx^*)-f_o(\underline\bx^{k}) &\le \frac{2\Lambda R^2+2LR^3/3}{k+3},\quad\text{for}\;k\ge2,
\end{align}
where
\begin{equation}
\Lambda = 
\left\{ 
\begin{array}{ll}
    \!\!\Lambda_q & \text{for Algorithm \ref{algorithm:QT}}
    \vspace{0.5em}\\
    \!\!\Lambda_t & \text{for Algorithm \ref{algorithm:GQT}}.\\
\end{array}
\right.
\end{equation}
\end{proposition}
\begin{IEEEproof}
See Appendix A.
\end{IEEEproof}

Because $0\le\Lambda_q\le\Lambda_t$, Algorithm \ref{algorithm:QT} converges faster than Algorithm \ref{algorithm:GQT} in iterations according to Proposition \ref{prop:MM_converge}. Notice that $\Lambda_q$ and $\Lambda_t$ in essence characterize how well their corresponding surrogate functions approximate the second-order profile of $f_o(\underline\bx)$. In the ideal case, the surrogate function and $f_o(\underline\bx)$ have exactly the same second-order profile so that $\Lambda=0$, then the objective-value error bound in Proposition \ref{prop:MM_converge} becomes
\begin{equation}
\label{cubic:bound}
    f_o(\underline\bx) - f_o(\underline\bx^k)\le\frac{L}{6}\|\underline\bx-\underline\bx^{k-1}\|^3_2,
\end{equation}
which also holds for the \emph{cubically regularized Newton's method} due to Nesterov as shown in \cite{Nesterov_book}. Equipped with the error bound \eqref{cubic:bound}, it immediately follows from Theorem 4.1.4 in \cite{Nesterov_book} that
\begin{align}
f_o(\underline\bx^*)-f_o(\underline\bx^{1}) &\le \frac{L R^3}{6}
    \label{ideal:error_bound1}\\
f_o(\underline\bx^*)-f_o(\underline\bx^{k}) &\le \frac{LR^3}{2(1+k/3)^2},\quad\text{for}\;k\ge2.
    \label{ideal:error_bound2}
\end{align} 
 
We now show that the extrapolated quadratic transform method in Algorithm \ref{algorithm:EQT} can achieve fairly close to the ideal case stated in \eqref{ideal:error_bound1} and \eqref{ideal:error_bound2}.
Since Algorithm \ref{algorithm:GQT} is a gradient projection method and Algorithm \ref{algorithm:EQT} accelerates it by Nesterov's extrapolation, we immediately obtain the following convergence rate from Proposition 6.2.1 of \cite{Bertsekas_book}.
\begin{proposition}[Convergence Rate of Algorithm \ref{algorithm:EQT}]
Suppose that the gradient of $f_o(\underline\bx)$ is $C$-Lipschitz continuous and let $\lambda_i^k = 1/(2C)$. Then Algorithm \ref{algorithm:EQT} yields
\begin{equation}
    f(\underline\bx^*) - f(\underline\bx) \le \frac{2C\cdot[f(\underline\bx^*) - f(\underline\bx^0)]}{(k+1)^2},\;\;\text{for}\; k\ge1.
\end{equation}
\end{proposition}
In summary, as compared to Algorithm \ref{algorithm:QT} and Algorithm \ref{algorithm:GQT} that both yield an objective-value error bound of $O(1/k)$, Algorithm \ref{algorithm:EQT} yields a smaller error bound of $O(1/k^2)$.

\section{Extension to Other FP Problems}
\label{sec:other_MFP}

The discussion thus far is limited to the sum-of-weighted-ratios problem in \eqref{prob:MFP}. Our goal here is to extend the above results to other FP problems. We begin with 
a recent discovery in \cite{ZP_MM+} that the joint application of the nonhomogeneous quadratic transform and the Lagrangian dual transform \cite{shen2018fractional2} can be interpreted as  a gradient projection method. We then generalize the above result for a general utility FP problem under certain conditions.

\subsection{Log-Ratio Case}

We first focus on the WSR problem and aim to rederive the main result in \cite{ZP_MM+} from an FP point of view. Consider $n$ interfering wireless links, each having $d$ transmit antennas and $\ell$ receive antennas. Let $\bx_i$ be the transmit beamformer of link $i$, let $\rho$ be the power constraint, and let $\mu_i$ be the rate weight of link $i$. Assume that one data stream is transmitted on each link. With $M_i(\underline\bx)$ interpreted as the SINR of link $i$, the WSR problem has the log-ratio form:
\begin{subequations}
\label{prob:example}
\begin{align}
  \underset{\underline\bx}{\text{maximize}} &\quad \sum^n_{i=1}\mu_i\log\big(1+M_i(\underline\bx)\big)\\
  \text{subject to} & \quad\, \|\bx_i\|^2_2\le\rho,\quad i=1,\ldots,n.
\end{align}
\end{subequations}
By the Lagrangian dual transform \cite{shen2018fractional2}, the above problem can be converted to the sum-of-weighted-ratios problem:
\begin{subequations}
\label{prob:example:h}
\begin{align}
  \underset{\underline\bx,\,\underline\gamma}{\text{maximize}} &\quad h(\underline\bx,\underline\gamma)\\
  \text{subject to} & \quad\, \|\bx_i\|^2_2\le\rho,\quad i=1,\ldots,n\\
  &\quad\, \gamma_i\in\mathbb R,\quad i=1,\ldots,n,
\end{align}
\end{subequations}
where
\begin{multline}
\label{example:h}
    h(\underline\bx,\underline\gamma) = \sum^n_{i=1}\Big[\mu_i(1+\gamma_i)\cdot \hat M_i(\underline\bx)\Big]\\
    +\sum^n_{i=1}\Big[\mu_i\log(1+\gamma_i)-\mu_i\gamma_i\Big]
\end{multline}
and
\begin{multline}
\label{hat M_i}
\hat M_i(\underline\bx) = \big(\bA_i\bx_i\big)^\hh\Bigg(\bA_i\bx_i\bx_i^\hh\bA_i^\hh+\sum^n_{j=1}\bB_{ij}\bx_{j}\bx_{j}^\hh\bB_{ij}^\hh\Bigg)^{-1}\\
 \big(\bA_i\bx_i\big).
\end{multline}
Observe that the optimization of $\underline\bx$ in \eqref{prob:example:h} amounts to a sum-of-weighted-ratios problem, so we can use the nonhomogeneous quadratic transform to further recast $h(\underline\bx,\underline\gamma)$ to
\begin{align}
\label{example:gt}
&f_t(\underline\bx,\underline\by,\underline\bz,\underline\gamma) =\sum^n_{i=1} \Big[ 2\Re\big\{\mu_i(1+\gamma_i)\bx^\hh_i\bA^\hh_i\by_i\notag\\
&\qquad\quad+\bx^\hh_i(\lambda_i\bI-\bD_i)\bz_i\big\}+\bz^\hh_i(\bD_i-\lambda_i\bI)\bz_i-\lambda_i\bx^\hh_i\bx_i\Big]\notag\\  &\qquad\quad+\sum^n_{i=1}\Big[\mu_i\log(1+\gamma_i)-\mu_i\gamma_i\Big],
\end{align}
where 
\begin{equation*}
    \bD_i = \mu_i(1+\gamma_i)\bA_i^\hh\by_i\by_i^\hh\bA_i
+\sum^n_{j=1}\mu_j(1+\gamma_j)\bB_{ji}^\hh\by_j\by^\hh_j\bB_{ji}
\end{equation*}
and
$$
\lambda_i \ge \lambda_{\max}(\bD_i).
$$
We optimize $(\underline\bx,\underline\by,\underline\bz,\underline\gamma)$ in \eqref{example:gt} iteratively. Again, by Lemma \ref{lemma:Taylor}, each $\bz_i$ is optimally determined as 
\begin{equation}
\label{example:z}    
\bz^\star_i = \bx_i.
\end{equation}
By completing the square for each $\by_i$ in \eqref{example:gt}, we obtain the optimal $\by_i$ as
\begin{align}
\label{example:y}
\by_i^\star &= \Bigg(\bA_i\bx_i\bx_i^\hh\bA_i^\hh+\sum^n_{i=1}\bB_{ij}\bx_j\bx^\hh_j\bB^\hh_{ij}\Bigg)^{-1}\bA_i\bx_i.   
\end{align}
After the above $\bz_i^\star$ and $\by_i^\star$ are plugged in \eqref{example:gt}, by solving $\partial  f_t/\partial \gamma_i=0$, each optimal $\gamma_i$ can be obtained as
\begin{align}
\label{example:gamma}
\gamma_i^\star &= 
M_i(\underline\bx).
\end{align}
Moreover, when $(\underline\by,\underline\bz,\underline\gamma)$ are all held fixed, we complete the square for each $\bx_i$ in $g_t(\underline\bx,\underline\by,\underline\bz,\underline\gamma)$ and solve for $\bx_i$ as
\begin{equation}
\label{example:x}
\bx^\star_i =
\left\{ 
\begin{array}{ll}
    \!\!\hat\bx_i & \text{if}\;\|\hat\bx_i\|^2_2\le\rho
    \vspace{0.5em}\\
    \!\!\big(\sqrt{\rho}/\|\hat\bx_i\|_2\big)\hat\bx_i & \text{otherwise},
\end{array}
\right.
\end{equation}
where
\begin{equation}
\label{example:x hat}
\hat\bx_i=\bz_i + \frac{1}{\lambda_i}\Big(\mu_i(1+\gamma_i)\bA^\hh_i\by_i-\bD_i\bz_i\Big).
\end{equation}
Putting \eqref{example:y}, \eqref{example:gamma}, \eqref{example:x}, and \eqref{example:x hat} together gives rise to the beamforming algorithm proposed in \cite{ZP_MM+}. Importantly, \cite{ZP_MM+} points out that its proposed beamforming algorithm amounts to a gradient projection method, as specified in the following proposition.

\begin{proposition}
\label{prop:ZP_MM+}
For the WSR problem \eqref{prob:example}, updating $\underline\bz$, $\underline\by$, $\underline\gamma$, and $\underline\bx$ by \eqref{example:z}, \eqref{example:y}, \eqref{example:gamma}, and \eqref{example:x}, respectively, in an iterative fashion is equivalent to the gradient projection:
\begin{equation*}
\bx^{k}_i =  \mathcal{P}_{\|\bx_i\|^2_2\le\rho}\bigg(\bx^{k-1}_i + \frac{1}{\lambda_i}\cdot\frac{\partial f_o(\underline\bx^{k-1})}{\partial \bx^c_i}\bigg),
\end{equation*}
where $f_o(\underline\bx)=\sum^n_{i=1}\mu_i\log\big(1+M_i(\underline\bx)\big)$ is the optimization objective of the WSR problem.
\end{proposition}

\begin{remark}
    For the log-ratio case, we treat each $\mu_j(1+\gamma_j)$ in \eqref{example:h} as the weight of the ratio $\hat M_i(\underline\bx)$, and then applying the conventional quadratic transform recovers the WMMSE algorithm \cite{cioffi_WMMSE,Shi_WMMSE}. Alternatively, we could have let  $\hat M_i(\underline\bx)$ absorb $\mu_i(1+\gamma_i)$ and then treat $\mu_i(1+\gamma_i)\hat M_i(\underline\bx)$ as the ratio term with weight one; \cite{shen2018fractional2} shows the above type of ratio term is more suited for the discrete FP solving. In fact, there are infinitely many ways of deciding which part is the ratio term and which part is the weight. The resulting quadratic transform method can be accelerated as in Algorithm \ref{algorithm:EQT}, regardless.
\end{remark}

\subsection{General Utility Case}

We now consider the following FP problem whose objective function is a general utility function of multiple ratios:
\begin{subequations}
\label{prob:utility}
\begin{align}
  \underset{\underline\bx}{\text{maximize}} &\quad \mathcal G\big(M_1(\underline\bx),\ldots,M_n(\underline\bx)\big)\\
  \text{subject to} & \quad \bx_i\in\mathcal X_i,\quad i=1,\ldots,n,
\end{align}
\end{subequations}
where $\mathcal G: \mathbb R^n\rightarrow\mathbb R$ is a differentiable function with $n$ ratio arguments. Our goal is to show that the gradient projection interpretation continues to hold for this general utility FP problem under certain condition; the previous log-ratio WSR problem case \cite{ZP_MM+} turns out to be its special case.

Assume that there exists a surrogate function $h(\underline\bx,\bg)$ of $\mathcal G\big(M_1(\underline\bx),\ldots,M_n(\underline\bx)\big)$ with the form of
\begin{equation}
\label{h}
h(\underline\bx,\bg) = \sum^n_{i=1}\Big[\alpha_i(\bg)\cdot \hat M_i(\underline\bx)\Big] + \beta(\bg),
\end{equation}
where $\bg$ is an auxiliary variable, $\alpha_i(\bg)\ge0$, $i=1,\ldots,n$, and $\beta(\bg)$ are differentiable scalar-valued functions of $\bg$, and
\begin{align}
\label{M_i hat}
 \hat M_i(\underline\bx) = \big(\hat\bA_i\bx_i\big)^\hh\Bigg(\sum^n_{j=1}\hat\bB_{ij}\bx_{j}\bx_{j}^\hh\hat\bB_{ij}^\hh\Bigg)^{-1}\big(\hat\bA_i\bx_i\big).
\end{align}
Note that $\hat M_i(\underline\bx)\ne M_i(\underline\bx)$ in general. Since $h(\underline\bx,\bg)$ is a surrogate function of $f_o(\underline\bx)$, we have
\begin{equation}  
\label{surrogate G}
\mathcal G\big(M_1(\underline\bx),\ldots,M_n(\underline\bx)\big) = \max_{\bg} h(\underline\bx,\bg).
\end{equation}
When $\bg$ is held fixed, $h(\underline\bx,\bg)$ can be recognized as a sum-of-weighted-ratios objective function of $\underline\bx$, so we further use the nonhomogeneous quadratic transform to reformulate $h(\underline\bx,\bg)$ as
\begin{multline}
\label{general utility f_t}
\!\!\!f_t(\underline\bx,\underline\by,\underline\bz,\bg) =  \sum^n_{i=1} \Big[ 2\Re\big\{\alpha_i(\bg)\bx^\hh_i\hat\bA^\hh_i\by_i+\bx^\hh_i(\lambda_i\bI-\bD_i)\bz_i\big\}\\
+\bz^\hh_i(\bD_i-\lambda_i\bI)\bz_i-\lambda_i\bx^\hh_i\bx_i\Big] + \beta(\bg),
\end{multline}
where
\begin{equation*}
\bD_i = \sum^n_{j=1}\alpha_j(\bg)\hat\bB_{ji}^\hh\by_j\by^\hh_j\hat\bB_{ji}
\end{equation*}
and
$$
\lambda_i \ge \lambda_{\max}(\bD_i).
$$
Again, we then optimize the variables $(\underline\bx,\underline\by,\underline\bz,\bg)$ iteratively in \eqref{general utility f_t}. According to Lemma \ref{lemma:Taylor}, the optimal update of $\bz_i$ is
\begin{equation}
\label{utility z}
    \bz^\star_i = \bx_i.
\end{equation}
By completing the square for each $\by_i$ in \eqref{general utility f_t}, the optimal $\by_i$ is obtained as
\begin{equation}
\label{utility y}
    \by^\star_i = \Bigg(\sum^n_{j=1}\hat\bB_{ij}\bx_j\bx^\hh_j\hat\bB^\hh_{ij}\Bigg)^{-1}\big(\hat\bA_i\bx_i\big).
\end{equation}
Subsequently, the optimal $\bg$ is given by
\begin{equation}
\label{utility g}
    \bg^\star = \arg\max_{\bg} h(\underline\bx,\bg).
\end{equation}
When the above variables are held fixed, we update each $\bx_i$ optimally as
\begin{align}
    \bx_i^\star 
    &= \mathcal{P}_{\mathcal X_i}\bigg(\bz_i + \frac{1}{\lambda_i}\big(\alpha_i(\bg)\hat\bA^\hh_i\by_i-\bD_i\bz_i\big)\bigg).
    \label{utility x}
\end{align}
We now show that the result of Proposition \ref{prop:ZP_MM+} carries over to the general utility case.

\begin{proposition}
\label{prop:general_gradient}
For problem \eqref{prob:utility}, suppose that there exists a surrogate function $h(\underline\bx,\bg)$ satisfying \eqref{h}, \eqref{M_i hat}, and \eqref{surrogate G}. Then updating $\underline\bz$, $\underline\by$, $\bg$, and $\underline\bx$ by \eqref{utility z}, \eqref{utility y}, \eqref{utility g}, and \eqref{utility x}, respectively, in an iterative fashion is equivalent to the gradient projection:
\begin{equation*}
\bx^{k}_i =  \mathcal{P}_{\bx_i\in\mathcal X_i}\bigg(\bx^{k-1}_i + \frac{1}{\lambda_i}\cdot\frac{\partial \mathcal G\big(M_1(\underline\bx),\ldots,M_n(\underline\bx)\big)}{\partial \bx^c_i}\bigg),
\end{equation*}
\end{proposition}
\begin{IEEEproof}
See Appendix B.
\end{IEEEproof}

Again, we could have adopted the extrapolated quadratic transform in place of the nonhomogeneous quadratic transform for the above method, and thereby accelerate the convergence.

\begin{figure*}[t]
 \setcounter{equation}{51}
\begin{align}
\label{matrix:fq}
f_q(\underline\bX,\underline\bY) &= \sum^n_{i=1}\Bigg[\omega_i\cdot\mathrm{tr}\Bigg(2\Re\{\bX^\hh_i\bA^\hh_i\bY_i\}-\bY^\hh_i\Bigg(\sum^n_{j=1}\bB_{ij}\bX_j\bX^\hh_j\bB^\hh_{ij}\Bigg)\bY_i\Bigg)\Bigg]
\end{align}
\hrule
\vspace{1em}
\setcounter{equation}{55}
\begin{equation}   
\label{matrix:ft}f_t(\underline\bX,\underline\bY,\underline\bZ) = \sum^n_{i=1}\Bigg[\mathrm{tr} \Big(2\Re\{\omega_i\bX^\hh_i\bA^\hh_i\bY_i+\bX^\hh_i(\lambda_i\bI-\bD_i)\bZ_i\}
    +\bZ^\hh_i(\bD_i-\lambda_i\bI)\bZ_i-\lambda_i\bX^\hh_i\bX_i\Big)\Bigg]
\end{equation}
\hrule
\end{figure*}

\section{Matrix Ratio Case}
\label{sec:matrix_var}

This section extends the preceding results to the generalized matrix ratios with the matrix variables $\bX_i\in\mathbb C^{d\times m}$ as in \eqref{matrix:ratio}. The sum-of-weighted-ratios FP problem in \eqref{prob:MFP} now becomes
\setcounter{equation}{50}
\begin{subequations}
\label{matrix:prob:MFP}
\begin{align}
  \underset{\underline\bX}{\text{maximize}} &\quad \sum^n_{i=1}\Big[\omega_i\cdot\mathrm{tr}\big(\bM_i(\underline\bX)\big)\Big]\\
  \text{subject to} & \quad\, \bX_i\in\mathcal X_i,\quad\text{for}\; i=1,\ldots,n.
\end{align}
\end{subequations}
The new objective function $f_q(\underline\bX,\underline\bY)$ by the quadratic transform is shown in \eqref{matrix:fq}, where an auxiliary variable $\bY_i\in\mathbb C^{d\times M}$ is introduced for each matrix ratio $\bM_i(\underline\bX)$.

Optimizing $\underline\bX$ and $\underline\bY$ alternatingly in $f_q(\underline\bX,\underline\bY)$ leads us to the matrix-ratio version of Algorithm \ref{algorithm:QT}, wherein
\setcounter{equation}{52}
\begin{equation}
\label{matrix:Y}
    \bY_i^\star = \Bigg(\sum^n_{i=1}\bB_{ij}\bX_j\bX^\hh_j\bB^\hh_{ij}\Bigg)^{-1}\big(\bA_i\bX_i\big)
\end{equation}
and
\begin{equation}
\label{matrix:X}
    \bX^\star_i = \arg\min_{\bX_i\in\mathcal X_i}\big\|\bD^{\frac12}_i\big(\bX_i-\omega_i\bD^{-1}_i\bA^\hh_i\bY_i\big)\big\|_F
\end{equation}
with
\begin{equation}
\label{matrix:Di}
    \bD_i = \sum^n_{j=1}\omega_j\bB_{ji}^\hh\bY_j\bY^\hh_j\bB_{ji}.
\end{equation}

We further extend the new objective function $f_t(\underline\bX,\underline\bY,\underline\bZ)$ of the nonhomogeneous quadratic transform as shown in \eqref{matrix:ft}, with an auxiliary variable $\bZ_i\in\mathbb C^{d\times m}$ introduced for each $\bD_i$. Again, we optimize the variables of $f_t(\underline\bX,\underline\bY,\underline\bZ)$ in an iterative fashion: $\underline\bZ$ is optimally updated to $\underline\bX$, $\underline\bY$ is optimally updated as in \eqref{matrix:Y}, and $\underline\bX$ is optimally updated as 
\begin{align*}
    \bX_i^\star 
    &= \mathcal{P}_{\mathcal X_i}\bigg(\bZ_i + \frac{1}{\lambda_i}\Big(\omega_i\bA^\hh_i\bY_i-\bD_i\bZ_i\Big)\bigg).
\end{align*}
Combining the above steps gives the matrix-ratio version of Algorithm \ref{algorithm:GQT}. Its connection with the gradient projection continues to hold:
\setcounter{equation}{56}
\begin{equation}
\bX^k = \mathcal{P}_{\mathcal X_i}\bigg(\bX^{k-1}_i + \frac{1}{\lambda^{k}_i}\cdot\frac{\partial f_o(\underline\bX^{k-1})}{\partial \bX_i^c}\bigg).
    \label{gradient_proj}
\end{equation}
Equipped with \eqref{gradient_proj}, Algorithm \ref{algorithm:EQT} can be immediately extended to the matrix ratio case as well.

\section{Two Application Cases}
\label{sec:applications}

\begin{figure}[t]
\centering
\includegraphics[width=4.6cm]{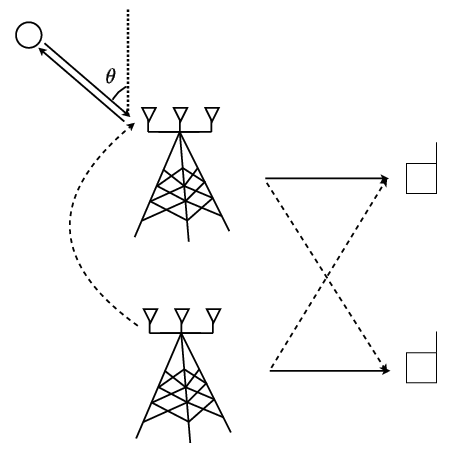}
\caption{Two BSs serve one downlink user each. One BS performs ISAC while the other BS only performs transmission; the aim of sensing is to recover the angle $\theta$. The dashed arrows represent the interference.}

\label{fig:ISAC}
\end{figure}

\subsection{Integrated Sensing and Communications (ISAC)}
Consider two base-stations (BSs) as depicted in Fig.~\ref{fig:ISAC}. BS 1 performs ISAC while BS 2 only performs downlink transmission. The two BSs have $M$ transmit antennas each, the two downlink users have $N$ antennas each, and BS 1 has $N_r$ radar receive antennas. Denote by $\bH_{ij}\in\mathbb C^{N\times M}$ the channel from BS $j$ to downlink user $i$, where $i,j\in\{1,2\}$, $\bG\in\mathbb C^{N_r\times M}$ the channel from BS 2 to the radar receiver at BS 1, $\sigma^2_i$ the background noise power at downlink user $i$, and $\sigma^2_r$ the background noise power at the radar antenna array of BS 1. Let $\bv_i\in\mathbb C^M$ be the transmit precoder at BS $i$ subject to the power constraint $P_{\max}$, i.e., $\|\bv_i\|^2_2\le P_{\max}$. 

\begin{figure*}[t]
\centering
\subfigure[convergence in iterations when $\omega_1=\omega_2=10^5$]{
\includegraphics[width=0.45\linewidth]{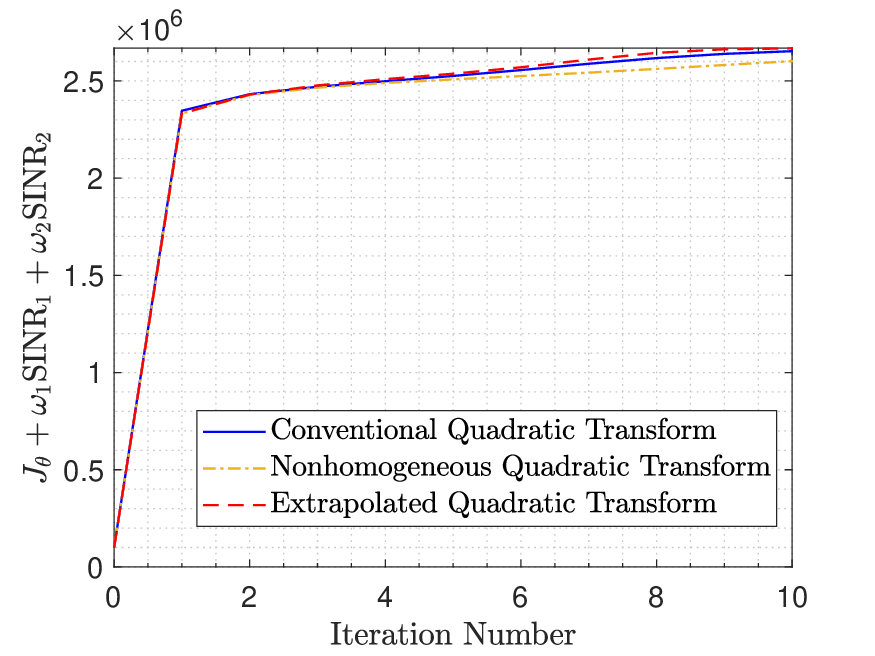}}
\hspace{1em}
\subfigure[convergence in time when $\omega_1=\omega_2=10^5$]{
\includegraphics[width=0.45\linewidth]{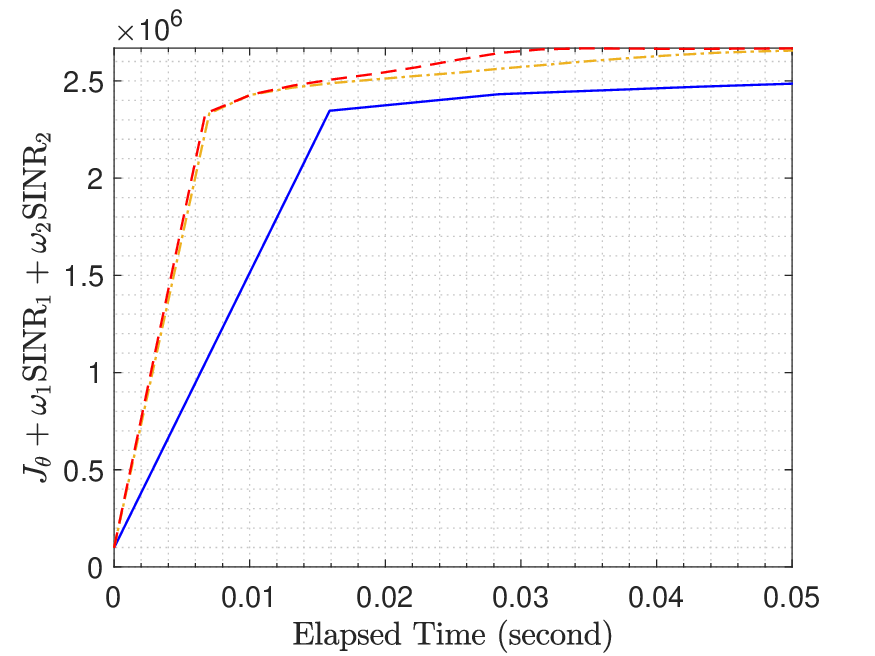}}\\
\subfigure[convergence in iterations when $\omega_1=\omega_2=10^9$]{
\includegraphics[width=0.45\linewidth]{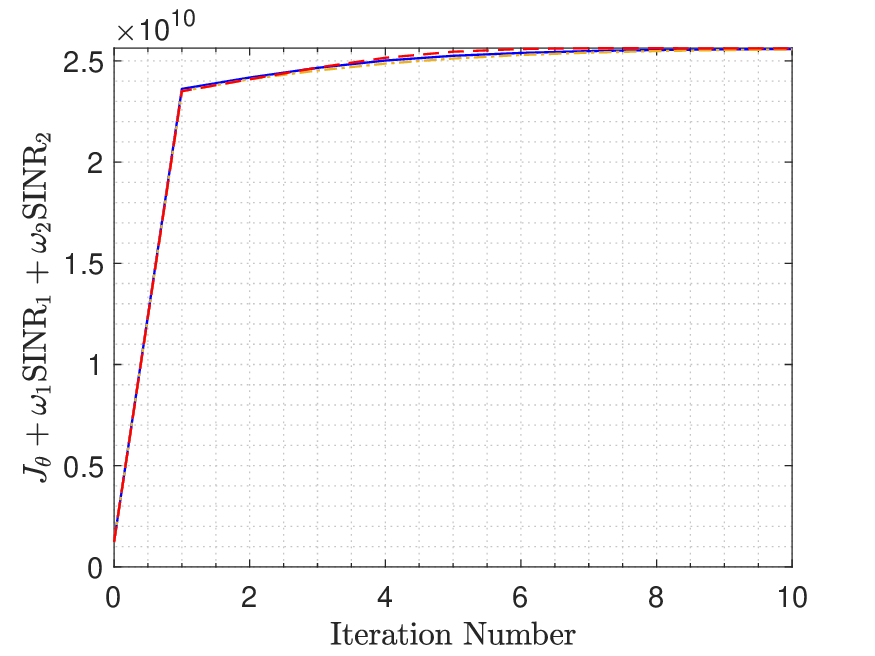}}
\hspace{1em}
\subfigure[{convergence in time when $\omega_1=\omega_2=10^9$}]{
\includegraphics[width=0.45\linewidth]{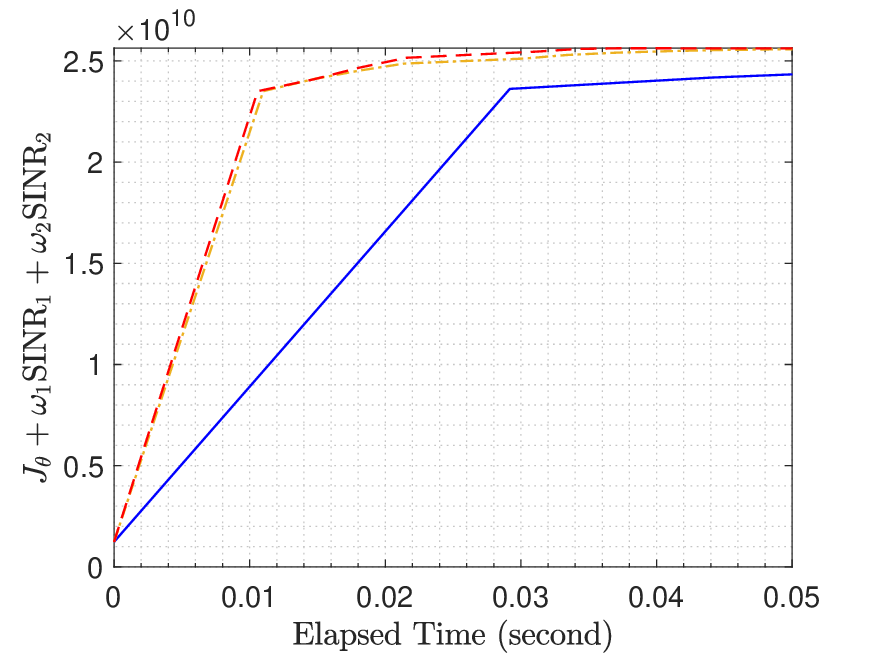}}
\caption{Maximizing a weighted sum of the Fisher information and the SINRs, $J_\theta+\omega_1\mathrm{SINR}_1+\omega_2\mathrm{SINR}_2$, for an ISAC system.}
\label{fig:ISAC:convergence}
\end{figure*}

Moreover, for BS 1, consider the transmit steering vector $\ba_{t}(\theta)\in\mathbb C^M$ and the receive steering vector $\ba_{r}(\theta)\in\mathbb C^{N_r}$, both dependent on the target angle $\theta$ as shown in Fig.~\ref{fig:ISAC}:
\begin{align*}
    \ba_{t}(\theta)&=\big[1,e^{-j\pi\sin\theta},\ldots,e^{-j\pi(M-1)\sin\theta}\big]^\top\\
    \ba_{r}(\theta)&=\big[1,e^{-j\pi\sin\theta},\ldots,e^{-j\pi(N_r-1)\sin\theta}\big]^\top.
\end{align*}
Thus, for the complex Gaussian symbol $\bs_i\sim\mathcal{CN}(\mathbf0,\bI)$ from BS $i\in\{1,2\}$, the received echo signal at BS 1 is given by
\begin{equation}
\label{eq:radar receive}
r=\xi\ba_{r}(\theta)\ba_{t}(\theta)^{\top}\bv_1\bs_{1}+\bG\bv_2\bs_2+\bz,
\end{equation}
where $\xi\in\mathbb{C}$ is the reflection coefficient, and $\bz\sim\mathcal{CN}(\mathbf 0,\sigma^2_r\bI)$ is the background noise at BS 1. Let $\hat{\bF}=\bG\bv_2\bs_2+\bz$ and define the interference-plus-noise covariance matrix to be
\begin{equation}
    \bQ =\mathbb E[\hat{\bF}\hat{\bF}^{\text{H}}] = \sigma^2_r\bI+\bG\bv_2\bv_2^\hh\bG^\hh.
\end{equation}
The Fisher information about the target angle $\theta$ in Fig.~\ref{fig:ISAC} is
\begin{equation}
    J_\theta =  \alpha\bv_1^\hh \dot{\bA}^\hh \bQ^{-1}\dot{\bA} \bv_1,
\end{equation}
where $\bA = \ba_{r}(\theta)\ba_{t}(\theta)^\top$,   $\dot{\bA}=\partial \bA/\partial \theta$, and $\alpha=2|\xi|^2$. The SINRs of the two downlink users are given by
\begin{align}
    \mathrm{SINR}_1 &= \bv_{1}^\hh\bH_{11}^\hh\Big(\sigma^2_1\bI+\bH_{12}\bv_{2}\bv_{2}^\hh\bH_{12}^\hh\Big)^{-1}\bH_{11}\bv_{1}\\
    \mathrm{SINR}_2 &= \bv_{2}^\hh\bH_{22}^\hh\Big(\sigma^2_2\bI+\bH_{21}\bv_{1}\bv_{1}^\hh\bH_{21}^\hh\Big)^{-1}\bH_{22}\bv_{2}.
\end{align}
We seek the optimal precoders $\underline\bv=\{\bv_1,\bv_2\}$ to maximize a linear combination of the Fisher information and the SINRs:
\begin{subequations}
\label{prob:ISAC}
\begin{align}
  \underset{\underline\bv}{\text{maximize}} &\quad J_\theta + \omega_1\mathrm{SINR}_1+\omega_2\mathrm{SINR}_2
  \label{prob:ISAC:obj}\\
  \text{subject to} & \quad \|\bv_{i}\|^2_2\le P_{\max},\;\;i\in\{1,2\},
  \label{prob:ISAC:constraint}
\end{align}
\end{subequations}
where $\omega_i>0$ reflects the priority of $\mathrm{SINR}_i$.

By the conventional quadratic transform, the original objective function can be recast to
\setcounter{equation}{63}
\begin{align}
&f_q(\underline\bv,\underline\by) = 2\Re\{\alpha\bv_1^\hh \dot{\bA}^\hh\by_r\} - \alpha\by_r^\hh\Big(\sigma^2_r\bI+\bG\bv_2\bv_2^\hh\bG^\hh\Big)\by_r\notag\\
&\;+2\Re\{\omega_1\bv_1^\hh\bH_{11}^\hh\by_{1}\}-\omega_1\by_{1}^\hh\Big(\sigma^2_1\bI+\bH_{12}\bv_{2}\bv_{2}^\hh\bH_{12}^\hh\Big)\by_{1}\notag\\
&\;+2\Re\{\omega_2\bv_2^\hh\bH_{22}^\hh\by_{2}\}- \omega_2\by_{2}^\hh\Big(\sigma^2_2\bI+\bH_{21}\bv_{1}\bv_{1}^\hh\bH_{21}^\hh\Big)\by_{2},
\end{align}
where the auxiliary variables $\by_r\in\mathbb C^{N_r}$, $\by_1\in\mathbb C^N$, and $\by_2\in\mathbb C^N$ are introduced for $J_\theta$, $\mathrm{SINR}_1$, and $\mathrm{SINR}_2$, respectively. We optimize the precoders $\underline\bv$ and the auxiliary variables $\underline\by=(\by_r,\by_1,\by_2)$ alternatingly as
\allowdisplaybreaks
\begin{equation*}
\underline\bv^0\rightarrow\cdots\rightarrow\underline\bv^{k-1} \rightarrow \underline\by^k\rightarrow \underline\bv^{k} \rightarrow \cdots.
\end{equation*}
For fixed $\underline\bv$, the optimal $\underline\by$ in $f_q(\underline\bv,\underline\by)$ is given by
\begin{subequations}
\begin{align}
\by_r^\star &= \Big(\sigma^2_r\bI+\bG\bv_2\bv_2^\hh\bG^\hh\Big)^{-1}\dot{\bA} \bv_1
\label{opt ys:ISAC}\\
\by_1^\star &= \Big(\sigma^2_1\bI+\bH_{12}\bv_{2}\bv_{2}^\hh\bH_{12}^\hh\Big)^{-1}\bH_{11}\bv_{1}
\label{opt y1:ISAC}\\
\by_2^\star &= \Big(\sigma^2_2\bI+\bH_{21}\bv_{1}\bv_{1}^\hh\bH_{21}^\hh\Big)^{-1}\bH_{22}\bv_{2}.
\label{opt y2:ISAC}
\end{align}
\end{subequations}
Next, we further find the optimal $\underline\bv$ in closed form as
\begin{subequations}
\begin{align}
    \bv_1^\star &= \Big(\eta_1\bI+\bD_1\Big)^{-1}\Big(\alpha\dot{\bA}^H\by_r+\omega_1\bH_{11}^\hh\by_{1}\Big)\\
    \bv_2^\star &= \Big(\eta_2\bI+\bD_2\Big)^{-1}\omega_2\bH_{22}^\hh\by_{2},
\end{align}
\end{subequations}
where
\begin{subequations}
\begin{align}
\bD_{1} &= \omega_2\bH_{21}^\hh\by_{2}\by_{2}^\hh\bH_{21}\\
\bD_2 &= \alpha\bG^\hh\by_r\by_r^\hh\bG+\omega_1\bH_{12}^\hh\by_{1}\by_{1}^\hh\bH_{12},
\end{align}
\end{subequations}
and the Lagrange multipliers $(\eta_1,\eta_2)$ for the power constraint are optimally determined as
\begin{equation}
    \label{eta:ISAC}
    \eta_{i}^\star = \min\big\{\eta_i\ge0:\|\bv_{i}\|^2_2\le P_{\max}\big\}.
\end{equation}
In practice, we may first try $\eta^\star_i=0$; if $\|\bv_i\|^2_2> P_{\max}$, then we tune $\eta^\star$ via bisection search to render $\|\bv_i\|^2_2 = P_{\max}$.

\begin{figure*}[t]
\setcounter{equation}{68}
\begin{multline}  
\label{ft:ISAC}
f_t(\underline\bv,\underline\by,\underline\bz) = 2\Re\Big\{\bv^\hh_1\Big[\alpha\dot{\bA}^\hh\by_r+ \omega_1\bH_{11}^\hh\by_1+(\lambda_{1}\bI-\bD_1)\bz_1\Big]+ \bv^\hh_2\Big[\omega_2\bH_{22}^\hh\by_2+(\lambda_2\bI-\bD_2)\bz_2\Big]\Big\}\\
+\bz^\hh_1(\bD_1-\lambda_{1}\bI)\bz_1+\bz^\hh_2(\bD_2-\lambda_2\bI)\bz_2-\lambda_{1}\|\bv_1\|^2_2-\lambda_2\|\bv_2\|^2_2-\alpha\sigma^2_r\|\by_r\|^2_2-\omega_1\sigma^2_1\|\by_1\|^2_2-\omega_2\sigma^2_2\|\by_2\|^2_2
\end{multline}
\hrule
\end{figure*}

Differing from the above conventional quadratic transform, the nonhomogeneous quadratic transform in Algorithm \ref{algorithm:EQT} recasts the original objective $f_o(\underline\bv)$ to $f_t(\underline\bv,\underline\by,\underline\bz)$ as shown in \eqref{ft:ISAC}. We optimize the variables in $f_t(\underline\bv,\underline\by,\underline\bz)$ iteratively:
\begin{equation*}
\underline\bv^0\rightarrow\cdots\rightarrow\underline\bv^{k-1} \rightarrow \underline\bz^k \rightarrow \underline\by^k \rightarrow \underline\bv^k \rightarrow \underline\bz^{k+1} \rightarrow \cdots.
\end{equation*}

\setcounter{equation}{63}
When $\underline\bv$ and $\underline\by$ are both held fixed, $\underline\bz$ is optimally updated as  $\bz_1^\star=\bv_1$, and $\bz_2^\star=\bv_2$. The optimal update of $\underline\by$ is the same as in \eqref{opt ys:ISAC}, \eqref{opt y1:ISAC}, and \eqref{opt y2:ISAC}. When $\underline\by$ and $\underline\bz$ are both held fixed, we first compute
\setcounter{equation}{69}
\begin{equation}
\hat\bv_1 = \bz_1+\frac{1}{\lambda_{1}}\Big(\alpha\dot{\bA}^\hh\by_r + \omega_1\bH_{11}^\hh\by_1-\bD_1\bz_1\Big)
\end{equation}
and
\begin{equation}
\hat\bv_2 = \bz_2+\frac{1}{\lambda_2}\Big(\omega_2\bH_{22}^\hh\by_2-\bD_2\bz_2\Big),
\end{equation}
and then update $\underline\bv$ optimally as
\begin{equation}
\label{}
\bv^\star_{i} =
\left\{ 
\begin{array}{ll}
    \!\!\hat\bv_{i}  &\text{if}\;\|\hat\bv_{i}\|^2_2\le P_{\max}
    \vspace{0.5em}\\
\!\!\big(\sqrt{P_{\max}}/\|\hat\bv_{i}\|_2\big)\hat\bv_{i}  &\text{otherwise}.
\end{array}
\right.
\end{equation}
We remark that other settings of ISAC have been considered in the literature with the Fisher information (or the Cram\'{e}r-Rao bound) written in a different form from our case, but the ISAC beamforming problem can still be handled by FP, e.g., as discussed in \cite{fan2018constant,guo2023bistatic1,guo2023bistatic}.

We now compare the various quadratic transform methods in simulations. Let $M=64$, $N=2$, $N_r = 72$, $\sigma_1^2=\sigma_2^2 = -80$ dBm, $\sigma_r^2 = -80$ dBm, and $P_{\max}=20$ dBm. The path loss (in dB) is computed as $32.6 + 36.7\log_{10}d$, where $d$ is the distance in meters; the position coordinates of BS 1, BS 2, user 1, user 2, and the sensed object are $(0,0)$, $(250,0)$, $(-10,100)$, $(350,100)$, and $(200,200)$, respectively, all in meters. The Rayleigh fading model is adopted. Algorithm \ref{algorithm:QT}, Algorithm \ref{algorithm:GQT}, and Algorithm \ref{algorithm:EQT} are tested. As shown in Fig.~\ref{fig:ISAC:convergence}(a), if the convergence is considered in terms of iterations, then all these algorithms yield almost the same convergence rate. The objective value is monotonically increasing with the iteration number by all these algorithms. If we instead evaluate convergence in terms of the elapsed time as displayed in Fig.~\ref{fig:ISAC:convergence}(b), then the proposed two accelerated methods, Algorithm \ref{algorithm:GQT} and Algorithm \ref{algorithm:EQT}, become much faster than Algorithm \ref{algorithm:QT}; the former two algorithms attain convergence after 0.02 seconds, whereas the latter algorithm still does not converge after 0.05 seconds. Algorithm \ref{algorithm:EQT} outperforms Algorithm \ref{algorithm:GQT}, but their gap is marginal. Moreover, as shown in Fig.~\ref{fig:ISAC:convergence}(c) and Fig.~\ref{fig:ISAC:convergence}(d), the above results still hold when the SINR weights are altered.

\subsection{Massive Multiple-Input Multiple-Output (MIMO)}
\label{subsec:MIMO}
\begin{figure}
\centering
{
\includegraphics[width=6.4cm]{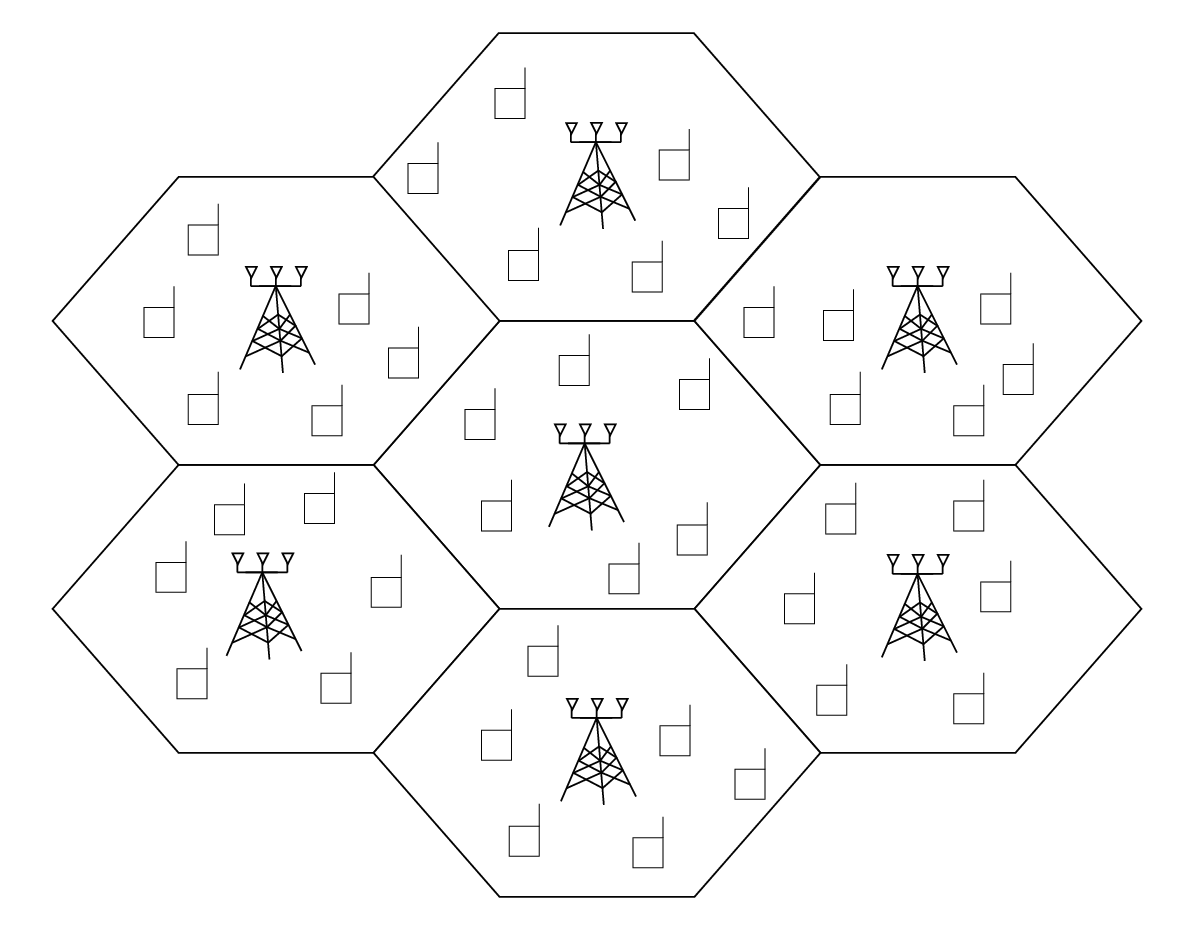}
\label{}
}
\caption{A 7-cell downlink wrapped-around massive MIMO network. We seek the optimal beamformers to maximize the sum of weighted rates across cells.}
\label{fig:massive MIMO}
\end{figure}

\begin{figure*}[t]
\centering
\subfigure[when $M=64$, $N=2$, $\sigma^2=-90$ dBm]{
\includegraphics[width=0.32\linewidth]{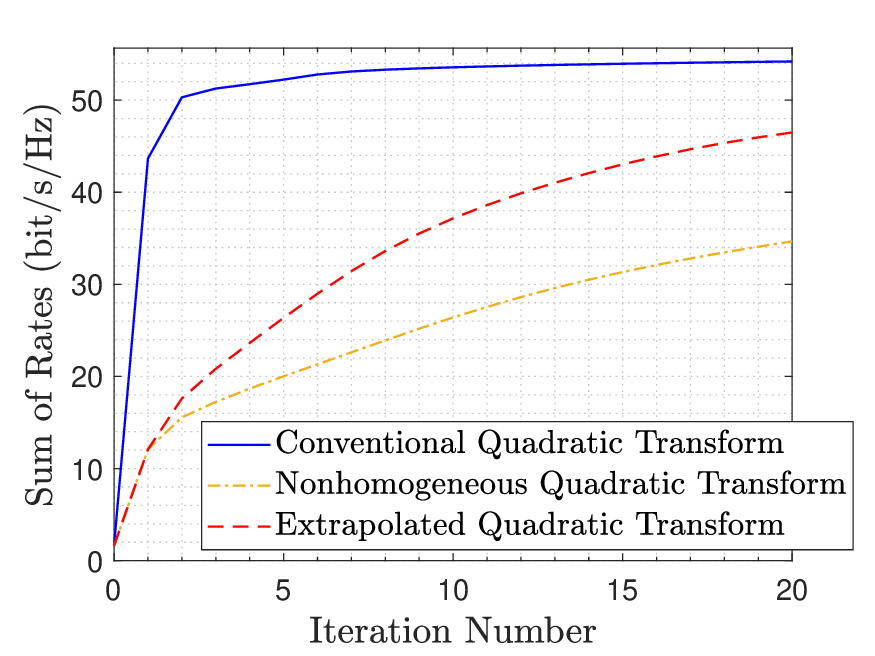}}
\subfigure[when $M=128$, $N=4$, $\sigma^2=-90$ dBm]{
\includegraphics[width=0.32\linewidth]{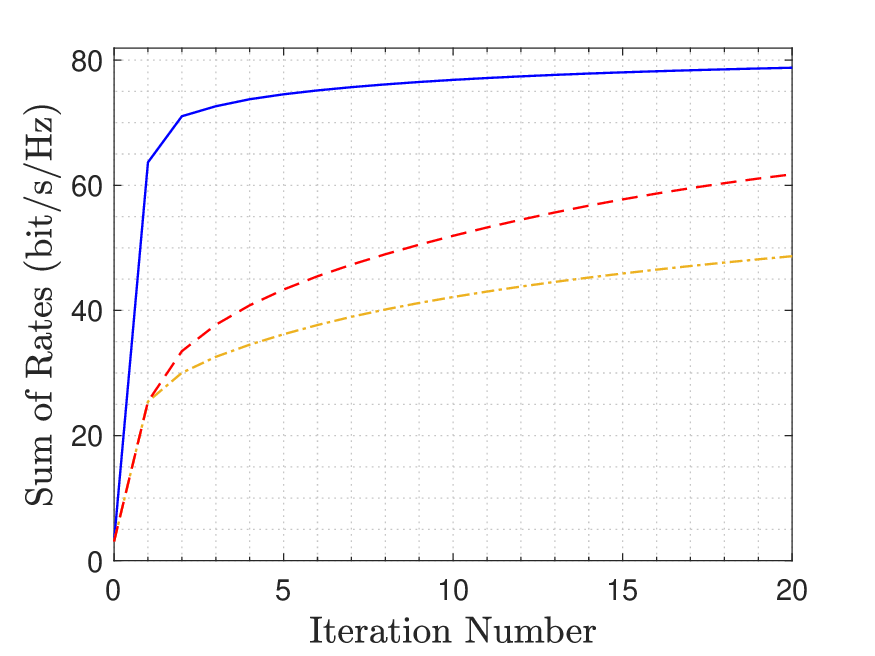}}
\subfigure[when $M=128$, $N=4$, $\sigma^2=-80$ dBm]{
\includegraphics[width=0.32\linewidth]{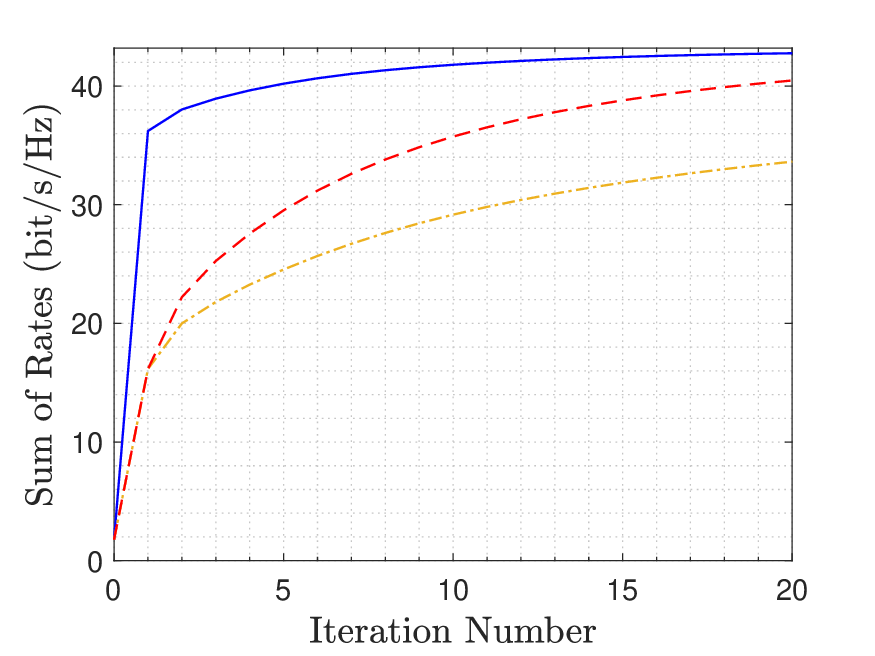}}
\\
\subfigure[when $M=64$, $N=2$, $\sigma^2=-90$ dBm]{
\includegraphics[width=0.32\linewidth]{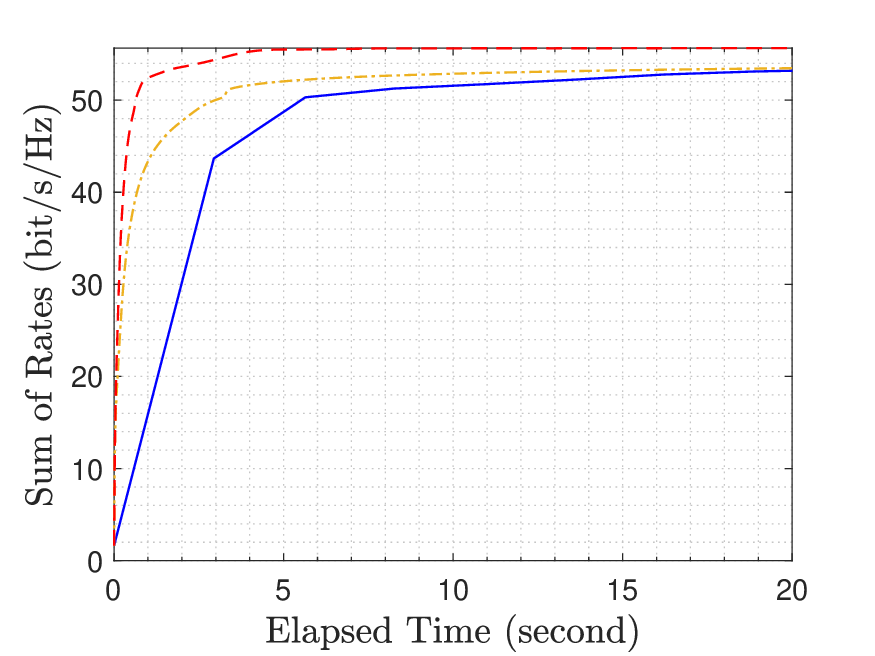}}
\subfigure[when $M=128$, $N=4$, $\sigma^2=-90$ dBm]{
\includegraphics[width=0.32\linewidth]{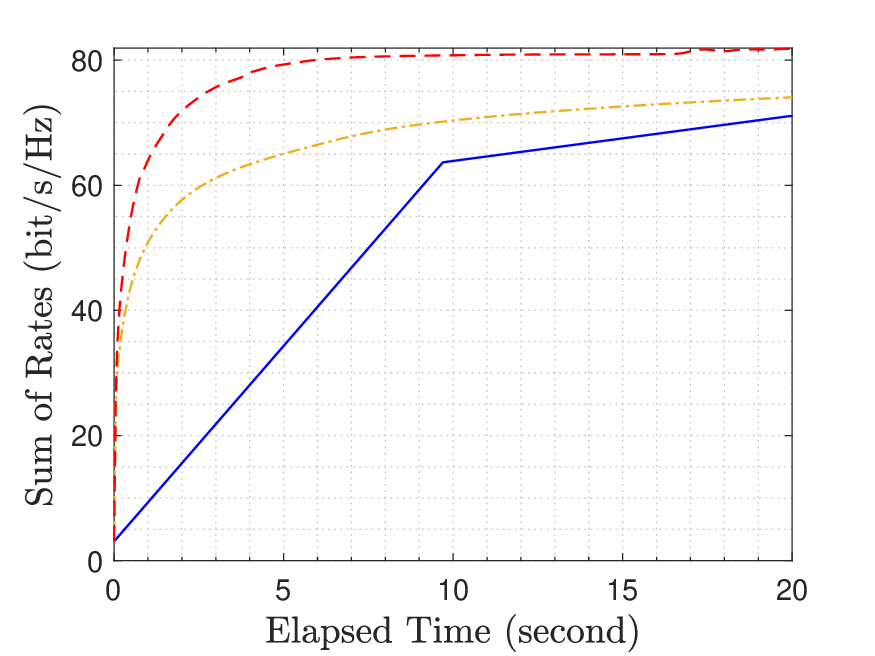}}
\subfigure[when $M=128$, $N=4$, $\sigma^2=-80$ dBm]{
\includegraphics[width=0.32\linewidth]{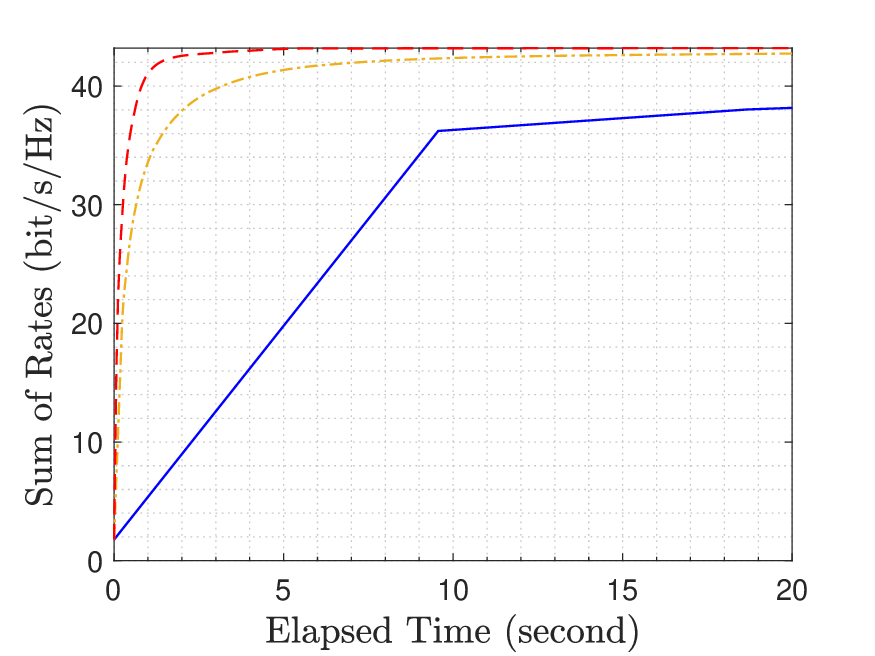}}
\caption{Maximizing the sum of rates in a multi-cell downlink massive MIMO network. On the top row the three figures show the convergence in iterations, while on the bottom row the three figures show the convergence in time.}
\label{fig:MIMO}
\end{figure*}

Consider a downlink network with $L$ cells as depicted in Fig.~\ref{fig:massive MIMO}. In each cell, the BS with $M$ antennas sends independent messages to $Q$ downlink user terminals simultaneously by spatial multiplexing; it shall be well understood that $Q\le M$. Assume also that each user terminal has $N$ receive antennas. In particular, $M\gg N$ under the massive MIMO setting.

Moreover, we use $\ell,i=1,\ldots,L$ to index the cells and the corresponding BSs, and use $q,j=1,\ldots,Q$ to index the users in each cell. Denote by $\bH_{\ell q,i}\in\mathbb C^{N\times M}$ the channel from BS $i$ to the $q$th user in cell $\ell$, denote by $\bv_{\ell q}\in\mathbb C^{M}$ the transmit precoder of BS $\ell$ for its $q$th associated user, and denote by $\sigma^2$ the background noise power. The SINR of the $q$th user in cell $\ell$, denoted by $\mathrm{SINR}_{\ell q}$, is computed\footnote{The SINR in \eqref{SINR:MIMO} can be achieved by using the MMSE receiver.} as
\begin{multline}
\label{SINR:MIMO}
\mathrm{SINR}_{\ell q} = \bv_{\ell q}^\hh\bH_{\ell q,\ell}^\hh\Bigg(\sigma^2\bI+\sum_{(i,j)\ne (\ell,q)}\bH_{\ell q,i}\bv_{ij}\bv_{ij}^\hh\bH_{\ell q,i}^\hh\Bigg)^{-1}\\
\bH_{\ell q,\ell}\bv_{\ell q}.
\end{multline}
Assigning a positive weight $\mu_{\ell q}>0$ for each user $q$ in cell $\ell$, we seek the optimal set of precoding vectors $\underline\bv=\{\bv_{\ell q}\}$ to maximize the weighted sum-of-rates throughout the network:
\begin{subequations}
\label{prob:MIMO}
\begin{align}
\underset{\underline\bv}{\text{maximize}} &\quad \sum^L_{\ell=1}\sum^Q_{q=1}\mu_{\ell q}\log\big(1+\mathrm{SINR}_{\ell q}\big)
\label{prob:MIMO:obj}\\
  \text{subject to} & \quad \sum^Q_{q=1}\|\bv_{\ell q}\|^2_2\le P_{\max},\;\text{for}\; \ell=1,\ldots,L,
  \label{prob:MIMO:constraint}
\end{align}
\end{subequations}
where the constraint \eqref{prob:MIMO:constraint} states that the total transmit power at each BS cannot exceed the power budget $P_{\max}$.

The traditional WMMSE method \cite{cioffi_WMMSE,Shi_WMMSE} addresses the above problem by performing the following iterative updates:
\begin{equation*}
\underline\bv^0\rightarrow\cdots\rightarrow\underline\bv^{k-1} \rightarrow \underline\by^k  \rightarrow \underline\gamma^k\rightarrow \underline\bv^{k} \rightarrow \cdots,
\end{equation*}
where the auxiliary variable $\underline\gamma$ is updated as
\begin{equation}
\label{t:MIMO}
\gamma^\star_{\ell q} = \mathrm{SINR}_{\ell q}
\end{equation}
for the current $\underline\bv$, and the auxiliary variable $\underline\by$ is updated as
\begin{align}
\label{y:MIMO}
\by^\star_{\ell q} &= \Bigg(\sigma^2\bI+\sum^L_{i=1}\sum^Q_{j=1}\bH_{\ell q,i}\bv_{ij}\bv^\hh_{ij}\bH^\hh_{\ell q,i}\Bigg)^{-1}\bH_{\ell q,\ell}\bv_{\ell q}.
\end{align}
With the auxiliary variables held fixed, the precoding vectors are optimally updated as
\begin{multline}
\label{xhat:WMMSE}
\bv_{\ell q}(\eta_{\ell}) = \Bigg(\eta_{\ell}\bI+\sum^L_{i=1}\sum^Q_{j=1}\mu_{ij}(1+\gamma_{ij})\bH^\hh_{ij,\ell}\by_{ij}\by^\hh_{ij}\bH_{ij,\ell}\Bigg)^{-1}\\
\mu_{\ell q}(1+\gamma_{\ell q})\bH^\hh_{\ell q,\ell}\by_{\ell q},
\end{multline}
where the Lagrange multiplier $\eta_\ell$ accounts for the power constraint at BS $\ell$ and is optimally determined as
\begin{equation}
    \label{eta:WMMSE}
    \eta_{\ell}^\star = \min\Bigg\{\eta\ge0:\sum^Q_{q=1}\|\bv_{\ell q}(\eta)\|^2_2\le P_{\max}\Bigg\}.
\end{equation}
In practice, the above $\eta_\ell$ can be obtained via bisection search. The WMMSE algorithm frequently requires inverting an $M\times M$ matrix, so it can be quite costly when $M\gg N$.

In contrast, the nonhomogeneous quadratic transform reformulates the objective function as
\begin{align}
&f_t(\underline\bv,\underline\by,\underline\bz,\underline\gamma) =\sum^L_{\ell=1}\sum^Q_{q=1}
\Big[ 2\Re\big\{\mu_{\ell q}(1+\gamma_{\ell q})\bv^\hh_{\ell q}\bH^\hh_{\ell q,\ell}\by_{\ell q}
\notag\\
&+\bv^\hh_{\ell q}(\lambda_{\ell}\bI-\bD_{\ell})\bz_{\ell q}\big\}
+\bz^\hh_{\ell q}(\bD_{\ell}-\lambda_{\ell}\bI)\bz_{\ell q}-\lambda_{\ell}\bv^\hh_{\ell q}\bv_{\ell q}\notag\\
&-\mu_{\ell q}(1+\gamma_{\ell q})\sigma^2\by_{\ell q}^\hh\by_{\ell q}
+\mu_{\ell q}\log(1+\gamma_{\ell q})-\mu_{\ell q}\gamma_{\ell q}\Big],
\label{example:chen}
\end{align}
for which the iterative updates are carried out as
\begin{equation*}
\underline\bv^0\rightarrow\cdots\rightarrow\underline\bv^{k-1} \rightarrow \underline\bz^k \rightarrow \underline\by^k  \rightarrow \underline\gamma^k\rightarrow \underline\bv^{k} \rightarrow \cdots,
\end{equation*}
where the auxiliary variable $\underline\bz$ is updated as $\bz_{\ell q} = \bv_{\ell q}$, and the other two auxiliary variables $\underline\by$ and $\underline\gamma$ are updated as in \eqref{y:MIMO} and \eqref{t:MIMO}, respectively. To update $\underline\bv$, we first compute
\begin{equation}
\label{massive MIMO v}
\hat\bv_{\ell q}=\bz_{\ell q} + \frac{1}{\lambda_\ell}\Big(\mu_{\ell q}(1+\gamma_{\ell q})\bH^\hh_{\ell q,\ell}\by_{\ell q}-\bD_\ell\bz_{\ell q}\Big),
\end{equation}
where
\begin{equation}
\bD_{\ell} = \sum^L_{i=1}\sum^Q_{j=1}\mu_{ij}(1+\gamma_{ij})\bH_{ij,\ell}^\hh\by_{ij}\by^\hh_{ij}\bH_{ij,\ell},
\end{equation}
and then optimally enforce the power constraint as
\begin{equation*}
\bv^\star_{\ell q} =
\left\{ 
\begin{array}{ll}
    \!\!\hat\bv_{\ell q}  &\text{if}\;\sum^Q_{j=1}\|\hat\bv_{\ell j}\|^2_2\le P_{\max}
    \vspace{0.5em}\\
\!\!\sqrt{\frac{{P_{\max}}}{\sum^Q_{j=1}\|\hat\bv_{\ell j}\|^2_2}} \hat\bv_{\ell q}&\text{otherwise}.
\end{array}
\right.
\end{equation*}

As opposed to the updating formula \eqref{xhat:WMMSE} of the WMMSE algorithm, the update of $\underline\bv$ in \eqref{massive MIMO v} no longer incurs any matrix inverse. Even though Algorithm \ref{algorithm:GQT} still requires computing matrix inverse for updating the auxiliary variable $\underline\by$ as in \eqref{y:MIMO}, the matrix size is just $N\times N$ with $N\ll M$ and thus can be neglected. Moreover, the above beamforming method for massive MIMO can be accelerated via extrapolation as in Algorithm \ref{algorithm:EQT}.

We now test the various quadratic transform methods for massive MIMO in a simulated 7-hexagonal-cell wrapped-around network as considered in \cite{shen2018fractional1}.   Within each cell, the BS is located at
the center and the $6$ downlink users are randomly placed. Each BS has $128$ antennas and each user has $4$ antennas. The BS-to-BS distance equals $0.8$ km. The maximum transmit power level at the BS side equals $20$ dBm, and the background noise power level equals $-90$ dBm. The downlink distance-dependent
path-loss is simulated by $128.1 + 37.6 \log_{10}d + \tau $ (in dB),
where $d$ represents the BS-to-user distance in km, and $\tau$ is a zero-mean Gaussian random variable with $8$ dB standard deviation for the shadowing effect. We consider sum rate maximization by setting all the weights to $1$. Again, Algorithm \ref{algorithm:QT}, Algorithm \ref{algorithm:GQT}, and Algorithm \ref{algorithm:EQT} are the competitors. As shown in Fig.~\ref{fig:MIMO}(a), Algorithm \ref{algorithm:QT} converges faster than the other two methods in terms of iterations; this result agrees with the former discussion below Proposition \ref{prop:MM_converge}. When it comes to the convergence evaluated by time, as shown in Fig.~\ref{fig:MIMO}(d), the two accelerated quadratic transform methods are much more efficient than the conventional method in Algorithm \ref{algorithm:QT}. In particular, observe that Algorithm \ref{algorithm:EQT} is also much faster than Algorithm \ref{algorithm:GQT}, as opposed to the ISAC scenario in Fig.~\ref{fig:ISAC:convergence}. There are two reasons. First, there are more matrix ratio terms in the massive MIMO problem case; second, the FP of massive MIMO has a more complicated structure (with ratios nested in logarithms). Moreover, as shown in Fig.~\ref{fig:MIMO}(b) and Fig.~\ref{fig:MIMO}(e), the advantage of Algorithm \ref{algorithm:EQT} over the other two methods becomes larger when more antennas are deployed. Nevertheless, when the noise level is raised, as shown in Fig.~\ref{fig:MIMO}(c) and Fig.~\ref{fig:MIMO}(d), the advantage of Algorithm \ref{algorithm:EQT} shrinks. The reason is that the FP-based coordination between the different beamformers becomes less important in the low-SNR regime; the maximum ratio transmission (MRT), i.e., aligning each beamformer with its corresponding channel, now tends to be a better choice \cite{bjornson2014optimal}.


\section{Conclusion}
\label{sec:conclusion}

This work significantly develops the existing theory and algorithm of the quadratic transform---which is a state-of-the-art tool for FP, with an emphasis on the applications for wireless communications. To start with, we establish a connection between the quadratic transform and the gradient projection, thus generalizing the result in \cite{ZP_MM+} about the WSR problem to a much wider range of FP problems. The above connection leads us to the nonhomogeneous quadratic transform that eliminates the matrix inverse operation from the FP solving. Next, we propose a novel idea of accelerating the quadratic transform by means of Nesterov's extrapolation \cite{Nesterov_book}. We then provide convergence rate analysis for the quadratic transform (including the conventional version and the accelerated ones). Since the quadratic transform encompasses the WMMSE algorithm \cite{cioffi_WMMSE,Shi_WMMSE} as a special case, the convergence rate analysis accounts for WMMSE too. To the best of our knowledge, this is the first work that gives a formal analysis as to how fast the quadratic transform or WMMSE converges. Moreover, we demonstrate the practical usefulness of the accelerated quadratic transform through two application cases, ISAC and massive MIMO, both of which are envisioned to be the key components of the next-generation wireless networks.


\section*{Appendix A\\Proof of Proposition \ref{prop:MM_converge}}

We focus on the convergence rate of Algorithm \ref{algorithm:QT}; the convergence rate of Algorithm \ref{algorithm:GQT} can be obtained similarly.
Lemma 1.2.4 in \cite{Nesterov_book} states that for any twice-differentiable function $f(\bx)$ with $L$-Lipschitz continuous gradient, we have
\begin{multline*}
    \bigg|f(\bx')-f(\bx)-\nabla f(\bx)^\hh(\bx'-\bx)\\
    -\frac{1}{2}(\bx'-\bx)^\hh\nabla^2f(\bx)(\bx'-\bx)\bigg|\le\frac{L}{6}\|\bx'-\bx\|^3_2
\end{multline*}
given any two feasible $\bx$ and $\bx'$. Applying the above lemma to the function $\delta_q(\underline\bx|\underline\bx^{k-1})$ and using the results in \eqref{delta:conditions} give
\begin{align}
&\frac{L}{6}\|\underline\bx-\underline\bx^{k-1}\|^3_2\notag\\
&\ge
\delta_q(\underline\bx|\underline\bx^{k-1})\notag\\
&\quad-\frac{1}{2}(\underline\bx-\underline\bx^{k-1})^\hh\nabla^2\delta_q(\underline\bx^{k-1}|\underline\bx^{k-1})(\underline\bx-\underline\bx^{k-1})\notag\\
&\ge \delta_q(\underline\bx|\underline\bx^{k-1})-\frac{\Lambda_q}{2}\|\underline\bx-\underline\bx^{k-1}\|^2_2\notag\\
&= f_o(\underline\bx)-f_q(\underline\bx,\underline\by^k)-\frac{\Lambda_q}{2}\|\underline\bx-\underline\bx^{k-1}\|^2_2\notag\\
&\overset{(a)}{\ge} f_o(\underline\bx)-f_q(\underline\bx^k,\underline\by^k)-\frac{\Lambda_q}{2}\|\underline\bx-\underline\bx^{k-1}\|^2_2\notag\\
&\overset{(b)}{\ge} f_o(\underline\bx)-f_q(\underline\bx^k,\underline\by^{k+1})-\frac{\Lambda_q}{2}\|\underline\bx-\underline\bx^{k-1}\|^2_2\notag\\
&\overset{(c)}{=} f_o(\underline\bx)-f_o(\underline\bx^{k})-\frac{\Lambda_q}{2}\|\underline\bx-\underline\bx^{k-1}\|^2_2,
\label{rate:L}
\end{align}
where step $(a)$ follows since $\underline\bx^k$ maximizes $f_q(\underline\bx,\underline\by)$ for the current $\underline\by=\underline\by^k$, step $(b)$ follows since $\underline\by^{k+1}$ maximizes $f_q(\underline\bx,\underline\by)$ for the current $\underline\bx=\underline\bx^k$, and step $(c)$ follows by the property of the surrogate function. Following Nesterov's proof technique in \cite{Nesterov_book}, we let
\begin{equation}
\label{rate:x}
    \underline\bx = \pi\underline\bx^*+(1-\pi)\underline\bx^{k-1},
\end{equation}
where $\pi\in[0,1]$. Then the concavity of $f_o(\underline\bx)$ on $\mathcal X$ gives
\begin{equation}
\label{rate:concave}
    f_o(\underline\bx) \le \pi f_o(\underline\bx^*) + (1-\pi) f_o(\underline\bx^{k-1}).
\end{equation}
Denote the gap in the objective value as
\begin{equation}
    v_{k} = f_o(\underline\bx^*)-f_o(\underline\bx^{k}).
\end{equation}
Substituting \eqref{rate:x} and \eqref{rate:concave} into \eqref{rate:L} gives rise to
\begin{align}
v_{k}
&\le (1-\pi) v_{k-1}+\frac{\pi^2\Lambda_q}{2}\|\underline\bx-\underline\bx^{k-1}\|^2_2\notag\\
&\qquad+\frac{\pi^3L}{6}\|\underline\bx-\underline\bx^{k-1}\|^3_2\notag\\
&\le (1-\pi) v_{k-1}+\pi^2\bigg(\frac{\Lambda_qR^2}{2}+\frac{LR^3}{6}\bigg),
\label{rate:v}
\end{align}
where the second inequality follows by \eqref{small_constraint} and $0\le\pi\le1$. The choice of $\pi$ depends on $k$. 

When $k=1$, we let $\pi=1$ in \eqref{rate:v} and obtain
\begin{equation}
\label{rate:v1}
    v_1\le \frac{\Lambda_qR^2}{2}+\frac{LR^3}{6}.
\end{equation}
When $k\ge2$, we let
\begin{equation}
\label{rate:pi}
    \pi = \frac{v_{k-1}}{\Lambda_qR^2+LR^3/3}.
\end{equation}
It can be shown by induction that the above $\pi$ is always feasible (i.e., $0\le\pi\le1$) for all $k\ge2$. Plugging \eqref{rate:pi} in \eqref{rate:v} yields
\begin{equation}
    v_k \le v_{k-1}\bigg(1-\frac{v_{k-1}}{2\Lambda_qR^2+2LR^3/3}\bigg),
\end{equation}
which can be further rewritten as
\allowdisplaybreaks
\begin{align}
\frac{1}{v_{k}} &\ge \frac{1}{v_{k-1}}\cdot\bigg(1-\frac{v_{k-1}}{2\Lambda_qR^2+2LR^3/3}\bigg)^{-1}\notag\\
&\ge \frac{1}{v_{k-1}}\cdot\bigg(1+\frac{v_{k-1}}{2\Lambda_qR^2+2LR^3/3}\bigg)\notag\\
&= \frac{1}{v_{k-1}} + \frac{1}{2\Lambda_qR^2+2LR^3/3},
\label{rate:1/v}
\end{align}
where the second inequality follows since $(1-a)^{-1}>1+a$ for any $0\le a\le 1$. The result of \eqref{rate:1/v} immediately gives
\begin{align}
\frac{1}{v_k}
&\ge \frac{1}{v_1} + \frac{k-1}{2\Lambda_qR^2+2LR^3/3}\notag\\
&\ge \frac{k+3}{2\Lambda_qR^2+2LR^3/3},
\end{align}
where the second inequality is due to \eqref{rate:v1}. The proof is then completed for Algorithm \ref{algorithm:QT}. The case of Algorithm \ref{algorithm:GQT} can be verified similarly.

\section*{Appendix B\\Proof of Proposition \ref{prop:general_gradient}}

Because the optimal value of $\bg$ depends on the current $\underline\bx$, it can be written as a function of $\underline\bx$, i.e.,
\begin{equation}
\label{t}
\bg^\star = \arg\max_{\bg}h(\underline\bx,\bg)\triangleq\mathcal T(\underline\bx).
\end{equation}
Computing $\bg$ in \eqref{t} is an unconstrained differentiable problem, so the optimal $\bg^\star$ must satisfy the first-order condition
\begin{equation}
\frac{d\beta(\bg^\star)}{d\bg}+\sum^n_{j=1}\bigg[\frac{d\alpha_j(\bg^\star)}{d\bg}\cdot \hat M_j(\underline\bx)\bigg] = 0.
\end{equation}
As a result, the partial derivative of $\mathcal G\big(M_1(\underline\bx),\ldots,M_n(\underline\bx)\big)$ with respect to each $\bx_i^c$ can be simplified into
\begin{align}
&\frac{\partial \mathcal G\big(M_1(\underline\bx),\ldots,M_n(\underline\bx)\big)}{\partial\bx^c_i}\notag\\
&= \frac{\partial h(\underline\bx,\mathcal T(\underline\bx))}{\partial\bx^c_i}\notag\\
&=
\sum^n_{j=1}\bigg[\bigg(\frac{\partial\mathcal T(\underline\bx)}{\partial\bx^c_i}\bigg)^\top\frac{d\alpha_j(\bg^\star)}{d\bg}\cdot\hat M_j(\underline\bx)+\alpha_j(\bg^\star)\cdot\frac{\partial \hat M_j(\underline\bx)}{\partial\bx^c_i}\bigg]\notag\\
&\qquad +\bigg(\frac{\partial\mathcal T(\underline\bx)}{\partial\bx^c_i}\bigg)^\top\cdot\frac{d\beta(\bg^\star)}{d\bg}\notag\\
&=\sum^n_{j=1}\bigg[\alpha_j(\bg^\star)\frac{\partial \hat M_j(\underline\bx)}{\partial\bx^c_i}\bigg].
\label{partial_go}
\end{align}
We now optimize the variables of $f_t(\underline\bx,\underline\by,\underline\bz,\bg)$ in \eqref{general utility f_t} iteratively as
\begin{equation*}
\underline\bx^0\rightarrow\cdots\rightarrow\underline\bx^{k-1} \rightarrow \underline\bz^k \rightarrow \underline\by^k  \rightarrow \bg^k\rightarrow \underline\bx^{k} \rightarrow \cdots.
\end{equation*}
The optimal update of $\bx^k_i$ is
\begin{align}
\bx^k_i 
&\overset{(a)}{=} \mathcal{P}_{\mathcal X_i}\bigg(\bx^{k-1}_i + \frac{1}{\lambda_i^k}\Big(\alpha_i(\bg^{k})\hat\bA^\hh_i\by^k_i-\bD^k_i\bx^{k-1}_i\Big)\bigg)\notag\\
&\overset{(b)}{=} \mathcal{P}_{\mathcal X_i}\Bigg(\bx_i^{k-1} + \frac{1}{\lambda_i^k}\sum^n_{j=1}\Bigg[\alpha_j(\bg^{k})\cdot\frac{\partial  \hat M_j(\underline\bx^{k-1})}{\partial \bx^c_i}\Bigg]\Bigg),
\end{align}
where step $(a)$ follows since each $\bz_i^{k}$ has been updated to $\bx_i^{k-1}$, and step $(b)$ follows by Lemma \ref{lemma:derivative}. Substituting \eqref{partial_go} in the above equation completes the proof.

\bibliographystyle{IEEEtran}
\bibliography{IEEEabrv,refs}

\newpage

\begin{IEEEbiographynophoto}{Kaiming Shen}
(Senior Member, IEEE) received the B.Eng. degree in information security and the B.Sc. degree in mathematics from Shanghai Jiao Tong University, China in 2011, and then the M.A.Sc. degree in electrical and computer engineering from the University of Toronto, Canada in 2013. After working at a tech startup in Ottawa for one year, he returned to the University of Toronto and received the Ph.D. degree in electrical and computer engineering in early 2020. Dr. Shen has been with the School of Science and Engineering at The Chinese University of Hong Kong (CUHK), Shenzhen, China as a tenure-track assistant professor since 2020. His research interests include optimization, wireless communications, information theory, and machine learning.
Dr. Shen received the IEEE Signal Processing Society Young Author Best Paper Award in 2021, the CUHK Teaching Achievement Award in 2023, and the Frontiers of Science Award at the International Congress of Basic Science in 2024. Dr. Shen currently serves as an Editor for IEEE Transactions on Wireless Communications.
\end{IEEEbiographynophoto}

\vskip 0pt plus -1fil

\begin{IEEEbiographynophoto}{Ziping Zhao}
(Member, IEEE) received the B.Eng. degree in electronics and information engineering (with highest honor) from Huazhong University of Science and Technology (HUST), Wuhan, China, in 2014, and the Ph.D. degree in Electronic and Computer Engineering from the Hong Kong University of Science and Technology (HKUST), Hong Kong, in 2019. He also has held several visiting research positions in University of Minnesota, Twins City, MN, USA and HKUST, Hong Kong. Since 2019, he has been an Assistant Professor with the School of Information Science and Technology, ShanghaiTech University, Shanghai, China. His research interests include optimization and statistics with applications in signal processing and machine learning. His work was the recipient of the Best Student Paper Award from IEEE ICASSP 2023, and were the Best Student Paper Award finalists from IEEE SAM 2020 and IEEE SPAWC 2021. He was the recipient of the Hong Kong PhD Fellowship.
\end{IEEEbiographynophoto}

\vskip 0pt plus -1fil

\begin{IEEEbiographynophoto}{Yannan Chen}
(Student Member, IEEE) received the B.E. degree in automation engineering and the M.E. degree in pattern recognition and intelligent systems from Xiamen University, Xiamen, China, in 2018 and 2021, respectively. He is currently pursuing his Ph.D. degree with the School of Science and Engineering at The Chinese University of Hong Kong (Shenzhen). His research interests include optimization, wireless communications, and machine learning.
\end{IEEEbiographynophoto}

\vskip 0pt plus -1fil

\begin{IEEEbiographynophoto}{Zepeng Zhang}
(Student Member, IEEE) received the B.E. degree in measuring and control technology and instruments from Wuhan University in 2020 and the M.S. degree in information and communication engineering from ShanghaiTech University in 2023. He is currently pursuing the Ph.D. degree in Electrical Engineering at \'Ecole Polytechnique F\'ed\'erale de Lausanne. His research interests include graph machine learning and optimization.
\end{IEEEbiographynophoto}

\vskip 0pt plus -1fil

\begin{IEEEbiographynophoto}{Hei Victor Cheng}
(Member, IEEE) received the B.Eng. degree in electronic engineering from Tsinghua University, Beijing, China, the M.Phil. degree in electronic and computer engineering from the Hong Kong University of Science and Technology, and the Ph.D. degree from the Department of Electrical Engineering, Link\"{o}ping University, Sweden. He worked as a Post-Doctoral Research Fellow at the University
of Toronto, Toronto, ON, Canada. He is now an Assistant Professor with the Department of Electrical and Computer Engineering, Aarhus University, Denmark. His current research interests include next generation wireless technologies, intelligent surfaces, and machine learning.
\end{IEEEbiographynophoto}

\end{document}